\documentclass[aps,prd,10pt,twocolumn,preprintnumbers,superscriptaddress,nofootinbib,letterpaper,longbibliography]{revtex4-2}

\usepackage{hyperref}
\usepackage{graphicx}
\usepackage{amsfonts,amsmath,amssymb,bm}
\usepackage{array}
\usepackage{xcolor}
\usepackage[utf8]{inputenc}
\usepackage[explicit]{titlesec}
\usepackage{orcidlink}
\usepackage{ulem}
\usepackage{hyperref}
\hypersetup{
  colorlinks  = true,
  urlcolor    = blue,
  linkcolor   = blue,
  citecolor   = red
}

\usepackage{enumitem} 

\def\Fermilab{Theory Division, Fermi National Accelerator Laboratory, Batavia, IL 60510, USA}

\newcommand{\ml}{m_{\ell}}
\newcommand{\dsigma}{\frac{d^2\sigma}{d\Emu d\costheta}}
\newcommand{\dsigmav}{\frac{d^2\sigma}{dv_1 dv_2}}
\newcommand{\pND}{p_\mathrm{ND}}
\newcommand{\pFD}{p_\mathrm{FD}}
\newcommand{\qND}{q_\mathrm{ND}}
\newcommand{\qFD}{q_\mathrm{FD}}
\newcommand{\pFDtilde}{\tilde{p}_\mathrm{FD}}
\newcommand{\qFDtilde}{\tilde{q}_\mathrm{FD}}
\newcommand{\costheta}{\cos\theta}
\newcommand{\Emu}{E_{\ell}}
\newcommand{\El}{E_{\ell}}

\graphicspath{{figures/}}

\begin{document}


\title{Machine Learning Neutrino-Nucleus Cross Sections}

\author{Daniel C. Hackett\orcidlink{0000-0001-6039-3801}}
\email{dhackett@fnal.gov}
\affiliation{\Fermilab} 

\author{Joshua Isaacson\orcidlink{0000-0001-6164-1707}}
\email{isaacs21@msu.edu}
\affiliation{\Fermilab}
\affiliation{Deparment of Physics and Astronomy,
Michigan State University, East Lansing, MI 48824}

\author{Shirley~Weishi~Li\orcidlink{0000-0002-2157-8982}}
\email{shirley.li@uci.edu}
\affiliation{Department of Physics and Astronomy, University of California, Irvine, CA 92697}

\author{Karla Tame-Narvaez\orcidlink{0000-0002-2249-9450}}
\email{karla@fnal.gov}
\affiliation{\Fermilab} 

\author{Michael L. Wagman\orcidlink{0000-0001-7670-1880}}
\email{mwagman@fnal.gov}
\affiliation{\Fermilab}

\date{December 20, 2024}


\begin{abstract}
Neutrino-nucleus scattering cross sections are critical theoretical inputs for long-baseline neutrino oscillation experiments. However, robust modeling of these cross sections remains challenging. 
For a simple but physically motivated toy model of the DUNE experiment, we demonstrate that an accurate neural-network model of the cross section---leveraging only Standard-Model symmetries---can be learned from near-detector data. We perform a neutrino oscillation analysis with simulated far-detector events, finding that oscillation analysis results enabled by our data-driven cross-section model approach the theoretical limit achievable with perfect prior knowledge of the cross section. 
We further quantify the effects of flux shape and detector resolution uncertainties as well as systematics from cross-section mismodeling.
This proof-of-principle study highlights the potential of future neutrino near-detector datasets and data-driven cross-section models.
\end{abstract}

\preprint{FERMILAB-PUB-24-0948-T, UCI-HEP-TR-2024-25}

\maketitle


\section{Introduction}

Neutrinos serve as an excellent probe of the Standard Model and what lies beyond. After decades of extensive effort, neutrino physics is now entering a precision era, with next-generation experiments aiming to measure mixing parameters to percent-level accuracy~\cite{JUNO:2015zny,Hyper-Kamiokande:2018ofw,DUNE:2020ypp}. Consequently, the precision required for relevant theoretical inputs has significantly increased. A prominent example is the neutrino-nucleus scattering cross section in the GeV range, which is critical as neutrino-nucleus scattering is the primary detection channel used in long-baseline accelerator-based neutrino experiments~\cite{NuSTEC:2017hzk,Ruso:2022qes,deGouvea:2022gut}.

The primary ingredients needed to constrain neutrino oscillation parameters are incident neutrino energy \mbox{\emph{distributions}}.
However, because neutrinos are not directly observed in detectors, one typically reconstructs the incident neutrino energy of each event from the measured daughter particles~\cite{Ankowski:2015jya,CLAS:2021neh,NOvA:2021nfi,T2K:2021xwb,T2K:2023smv}. 
This reconstruction process relies on exclusive differential cross sections~\cite{Ankowski:2015kya,Friedland:2018vry,Friedland:2020cdp,CLAS:2021neh,MicroBooNE:2021tya,MicroBooNE:2024hun}; for example, accurate modeling of the energy of neutrons, which detectors often miss, is vital for accurately reconstructing the neutrino energy.
Therefore, cross-section models encapsulated in event generators are extensively utilized in neutrino experiments~\cite{Andreopoulos:2009rq,Buss:2011mx,Golan:2012rfa,MINERvA:2013zvz,Hayato:2021heg,Isaacson:2022cwh,MicroBooNE:2024zkh}. 

A first-principles calculation of neutrino-nucleus scattering cross sections proves to be a significant challenge. The nuclear materials used in neutrino experiments, such as carbon, oxygen, and argon, have complex internal structures. At low energies, they can be modeled as collections of protons and neutrons described by chiral effective field theory (EFT). At high energies, they can be accurately approximated as collections of quarks and gluons with interactions described by perturbative QCD. However, at medium energies of a few GeV, which coincide with the range of accelerator neutrino beam energies, constructing a systematically improvable EFT for nuclear physics remains difficult~\cite{Beane:2000fx,Epelbaum:2008ga,Kaplan:2019znu,Hammer:2019poc,vanKolck:2020llt,Epelbaum:2022cyo}. 

To address the challenges of cross-section modeling and other systematic uncertainties, oscillation experiments employ near detectors. By placing a detector close to the beam source---before oscillations are expected to occur---experiments can use near-detector (ND) events as validation tools for event generators. 
In a process called ND tuning, experiments use discrepancies between generator predictions and measured spectra to adjust generator models before using them to analyze far-detector (FD) samples~\cite{NOvA:2020rbg,NOvA:2021nfi,T2K:2021xwb,T2K:2023smv}. 
However, the accuracy of tuned cross sections relies on the validity of their underlying physics models and affects how well they can extrapolate from near- to far-detector kinematics. Significant cross-section uncertainties can still enter oscillation analyses after ND tuning~\cite{MINERvA:2019kfr,NOvA:2020rbg,NOvA:2021nfi,T2K:2021xwb,Coyle:2022bwa,T2K:2023smv}.

\begin{figure*}[t]
    \includegraphics[width=0.6\linewidth]{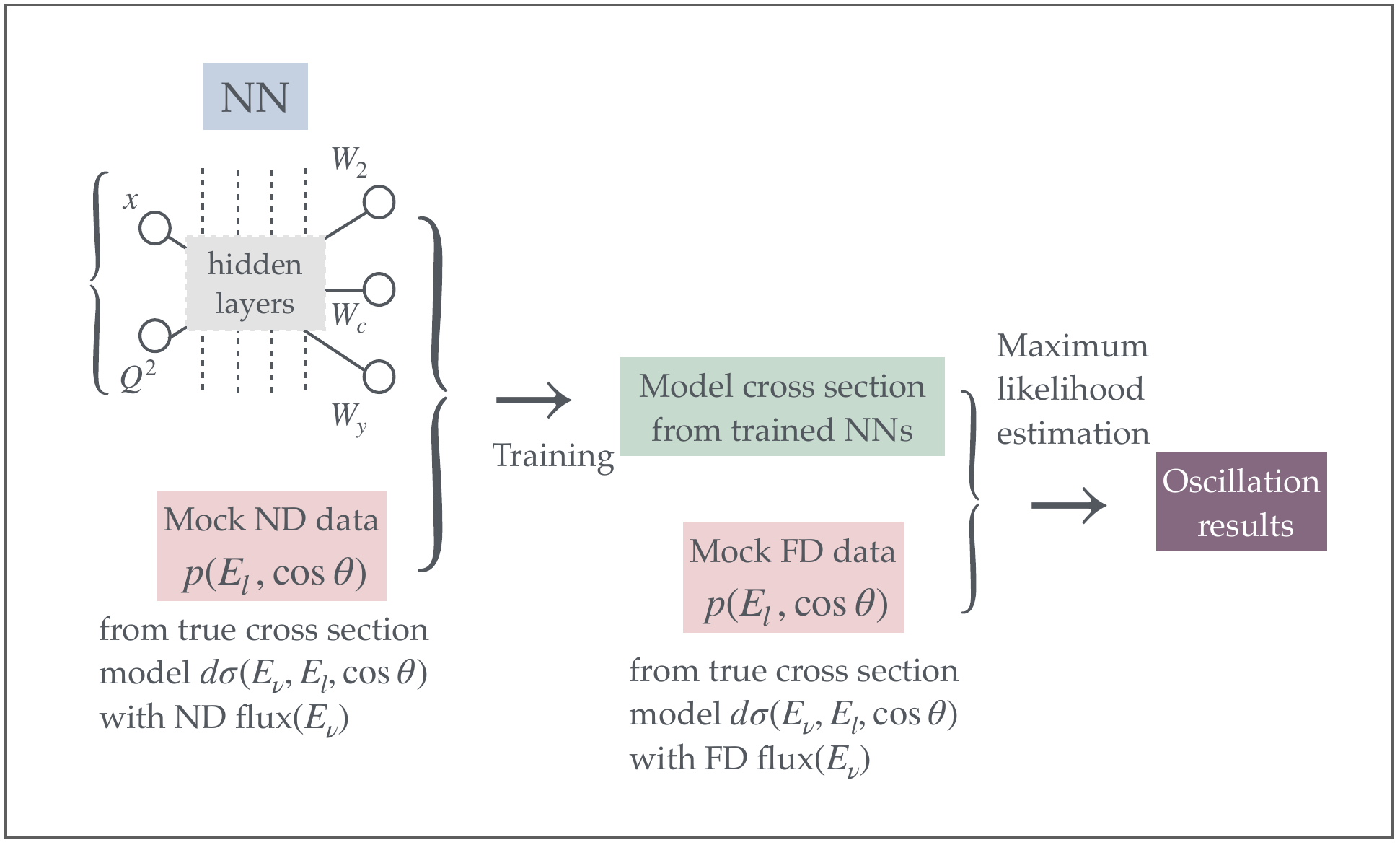}
    \caption{
        Illustration of our analysis procedure as applied in the context of our closure test. In a real application, the mock ND and FD data would be replaced by experimental data.
    }
    \label{fig:illustrate}
\end{figure*}

In this work, we explore an alternative approach to oscillation analysis using machine learning (ML).
To establish its viability, we consider only inclusive data in this initial exploration.
Figure~\ref{fig:illustrate} illustrates our analysis procedure. We construct a cross-section model using a neural network (NN) trained on mock ND data, specifically the outgoing muon energy $\Emu$ and angle $\costheta$. 
We then apply our cross-section model to determine oscillation parameters by optimizing agreement between mock FD data and predicted $(\Emu, \costheta)$ distributions. There is no event-by-event neutrino energy reconstruction in our approach; only distributions of neutrino energies and $(\Emu, \costheta)$ enter both our cross-section model training and our subsequent neutrino oscillation analysis.
The only theoretical assumption in our approach is that the inclusive neutrino-nucleus cross section can be parameterized by structure functions, which follows directly from Standard-Model symmetries.
Previous work has demonstrated that NN parameterizations can be used to accurately constrain one-dimensional parton distribution functions (PDFs) for both nucleons~\cite{Ball:2011uy} and nuclei~\cite{AbdulKhalek:2019mzd}, and more recently to model lepton-nucleus cross sections using two-dimensional structure/response functions such as those considered here~\cite{Forte:2002fg,Candido:2023utz,Sobczyk:2024ajv}. 

Our new approach is not meant to replace but rather complement the traditional one in several key aspects.
Our cross-section model is fully data-driven: rather than using ND data to jointly fit parameters of pre-defined flux and cross-section models---a process which can absorb mismodeling through unphysical parameter shifts---we directly learn a non-parametric representation of the cross-section from ND data. This learning process assumes the flux model can be constrained independently using e.g., neutrino-electron scattering data~\cite{Marshall:2019vdy}.
This ensures our model fully exploits the power of ND samples—incredible statistics and small detector systematics. Our method also offers the flexibility of adding layers of theoretical assumptions, e.g., relations between nuclear structure functions. Conversely, our method only applies to oscillation measurements and not to general new physics searches at the ND, which is an essential component of the accelerator neutrino program~\cite{T2K:2017ega,Machado:2019oxb,Altmannshofer:2019zhy,deGouvea:2019wav,Berryman:2019dme,Ellis:2020ehi,Ellis:2020hus,NOvA:2021smv,ArgoNeuT:2022mrm,MicroBooNE:2023eef,MicroBooNE:2023gmv,Coloma:2023oxx}.

To validate our approach in this proof-of-principle study, we conduct a closure test using a toy cross-section model with known structure functions. 
This allows us to directly assess how well our model learns the true cross section and
how this affects its ability to describe near- and far-detector flux-averaged cross sections. This closure test is a prerequisite to future studies that will apply the same approach to data, or to event generators, which will also test whether their underlying physics models admit decomposition into structure functions.
We also adopt several further simplifications that can all be relaxed in future studies. First, we use only the outgoing lepton information, specifically $\Emu$ and $\costheta$, and ignore any hadronic particles. Second, we consider only the oscillation channel $P(\nu_\mu \rightarrow \nu_\mu)$ and disregard all other channels. 
We begin by considering the idealized case of infinite ND statistics and without any detector effects such as energy resolution, but subsequently incorporate their effects as systematic uncertainties.


\section{Neutrino-nucleus scattering theory}

Consider charged-current scattering of a neutrino with initial energy $E_{\nu}$ on a nucleus into a final state consisting of a charged lepton with energy $\El$ and a hadronic remnant. The inclusive cross section can be parameterized in terms of a set of five structure functions~\cite{Albright:1974ts,Paschos:2001np,Kretzer:2002fr} as
 \begin{equation}
   \begin{split}
     &\frac{d^2 \sigma^{(\nu A)}}{d\El d\costheta}(E_\nu) = \frac{|V_{ud}|^2  G_F^2 \El}{\pi}   \sqrt{1 - \frac{\ml^2}{\El^2}}  \\
     &\hspace{25pt} \times \left\lbrace  \frac{E_{\nu}}{M_A} \left( 1 - y - \frac{Q^2 + \ml^2}{4 E_{\nu}^2} \right) W_2(x,Q^2)  \right. \\
     &\hspace{35pt} + \tilde{y} W_1(x,Q^2) + \left( 1 - \frac{\tilde{y}}{2} \right) W_3(x,Q^2)  \\
     &\hspace{35pt} \left.  - \left( \frac{\ml^2}{Q^2} \right) \left[ 2 W_5(x,Q^2) - \tilde{y}  W_4(x,Q^2) \right]  \vphantom{\frac{M_A xy}{2 E_{\nu}}} \right\rbrace ,
   \end{split} \label{eq:nuAd2}
\end{equation}
where $\theta$ is the lepton scattering angle, $\ml$ the charged lepton mass, $M_A$ the nuclear mass, $Q^2$ the four-momentum transfer squared, $x$ is Bjorken $x$, $y = Q^2 / (2 M_A E_{\nu} x)$ the inelasticity, and $\tilde{y} \equiv y(1 + \ml^2/Q^2)$.
The nuclear structure functions $W_i(x,Q^2)$ are defined from a Lorentz decomposition of $\bigl< A | J_{\mu}^\dagger J_{\nu} | A \bigr>$ where $J_{\mu} = \bar{u} \gamma_\mu (1-\gamma_5)d$ is an electroweak current and $\bigl | A \bigr>$ is the nuclear ground state.
Higher-order electroweak corrections and $\mathcal{O}(Q^2/m_W^2)$ effects are neglected here and throughout; see Refs.~\cite{Tomalak:2021hec,Tomalak:2022xup,Afanasev:2023gev} for discussion.
Factors of $x$ and $Q^2$ have been absorbed into the $W_i$ to remove zeros and poles from kinematic prefactors, which facilitates NN fitting.
They are related to the $F_i$ in Ref.~\cite{Kretzer:2002fr} by $W_i = x F_i$ for $i \in \{1, 3, 4, 5 \}$ and $W_2 = (2 x M_A^2/Q^2)  F_2$.
Contributions
from $W_4$ and $W_5$ are suppressed by $\ml^2 / Q^2$, which can reach 1--10\% for $\sim 1$ GeV muon neutrinos and are therefore relevant for DUNE's cross-section uncertainty targets.
Global fits of the structure functions have been studied in Ref.~\cite{Candido:2023utz}.

The essential feature of Eq.~\eqref{eq:nuAd2} is that the cross section depends on three independent kinematic variables, e.g., $(E_\nu, E_\ell, \costheta)$.
Inferring a three-dimensional function from the $E_\nu$-averaged two-dimensional distribution of $(E_\ell, \costheta)$ accessible in the ND is an ill-posed problem, as demonstrated in Appendix~\ref{sec:3d_bad} below.
The benefit of the structure function parameterization is that the $W_i$ depend on only two independent kinematic variables, $x$ and $Q^2$.
It is therefore possible to learn structure functions from ND data with some $(E_{\nu},x,Q^2)$ distribution and use them to analyze FD data, as long as the ND and FD marginal distributions over $(x,Q^2)$ are similar.
For DUNE, the ND and FD $(x,Q^2)$ distributions are expected to strongly overlap; neutrino oscillations will primarily redistribute events within the same kinematic region.
This is the key physics ingredient enabling our data-driven cross-section model and oscillation analysis.

In this work, we only consider the muon neutrino charged-current channel at both the ND and FD. Without multiple distinct lepton masses, two exact degeneracies arise between the structure functions, and the cross-section can be parameterized as
\begin{equation}
   \begin{aligned}
     \frac{d^2 \sigma^{(\nu A)}}{d\Emu d\costheta} &= \frac{|V_{ud}|^2  G_F^2 \Emu}{\pi}   \sqrt{1 - \frac{\ml^2}{\Emu^2}} 
     \left\lbrace  \frac{E_{\nu}}{M_A}  W_2(x,Q^2) 
     \right. \\
     &\left. 
     +  W_c(x,Q^2;\ml^2)
     + \tilde{y}  W_y(x,Q^2;\ml^2) \vphantom{\frac{M_A xy}{2 E_{\nu}}} \right\rbrace ,
   \end{aligned} \label{eq:nuAd2-unique}
\end{equation}
where $W_c = W_3 - Q^2 / (2 x M_A^2) W_2 - 2(\ml^2/Q^2)W_5$ and $W_y = W_1 - (x/2)W_2 - W_3/2 + (\ml^2/Q^2)W_4$. 
While not made explicit in the notation, we emphasize that the $W_i$ differ nontrivially between different nuclei.


\section{Closure test setup}

The fundamental question we seek to address is whether the cross section can be learned well enough to extract oscillation parameters.
Answering it with a closure test requires a fully known toy model of the physics of interest.
To this end, we define a set of structure functions $W_i$, a ND flux $\Phi_\mathrm{ND}$, and a FD flux $\Phi_\mathrm{FD}$, all as explicit functions that can be evaluated for any kinematics.
We describe these quantities as ``true'' or ``truth'' in the setting of the toy model.

For the structure functions, we take the leading order prediction from the quark-parton model~\cite{Bjorken:1969ja},
\begin{align}
  W_2 &= \frac{4x^2 M_A^2}{A Q^2} (\bar{u}+d+\bar{c}+s)\,, \\
    W_3 &= 2x(d-\bar{u}+s-\bar{c})\,,
\end{align}
with $W_1$ obtained using the Callan-Gross relation ($2x W_1= \frac{A Q^2}{2 M_A^2} W_2$)~\cite{Callan:1969uq}, and $W_4,W_5$ given by the tree-level relation from Ref.~\cite{Albright:1974ts} ($2x W_5= \frac{AQ^2}{2 M_A^2}  W_2,\, W_4=0$).
We choose the CT18NNLO PDFs~\cite{Hou:2019efy} for $\bar{u}$, $d$, $\bar{c}$, $s$, evaluated using LHAPDF6~\cite{Buckley:2014ana} and extrapolated outside the grid using the method of the MSTW collaboration~\cite{Martin:2009iq}.
When converting from nucleon structure functions to argon structure functions, the scaling discussed in Ref.~\cite{Ruiz:2023ozv} is applied.
Evaluating Eq.~\eqref{eq:nuAd2} with these $W_i$ defines the toy-model cross section.

We take the DUNE ND $\nu_\mu$ flux for the neutrino run-mode from Refs.~\cite{FieldsDune, DUNE:2020ypp}, linearly interpolated over $0 \leq E_{\nu_\mu} \leq 10~\mathrm{GeV}$ and defined as zero elsewhere. 
For the FD flux, we compute oscillation probabilities for a baseline of 1300 km, with truth parameters taken from the NuFit-6.0 fit~\cite{Esteban:2024eli} using the normal ordering: $\sin^2\theta_{23}=0.561$, $\sin^2\theta_{12}=0.307$, $\sin^2\theta_{13}=0.02195$, $\Delta m_{21}^2 = 7.49\times 10^{-5}$ eV${}^2$, $\Delta m_{31}^2 = 2.534\times 10^{-3}$~eV${}^2$, and $\delta_\text{CP} = 177^\circ$. 
The oscillations are calculated, including matter effects, using the NuFast package~\cite{Denton:2024pzc}.

The analysis involves two distinct statistical inference problems: learning the cross section at the ND, and extracting oscillation parameters at the FD.
We must therefore frame the problem in statistical language.
The product of a cross section and flux, $\dsigma \Phi$, defines a three-dimensional probability density of events $(E_\nu, \Emu, \costheta)$ after normalization.
However, without $E_\nu$ reconstruction, we have access to only $(\Emu, \costheta)$ for each event.
All available information is thus encoded by two-dimensional marginal densities of the form
\begin{equation}
\label{eq:marginalize}
    p(\Emu, \costheta) = \frac{ 
        \int dE_\nu \, \dsigma(E_\nu) \, \Phi(E_\nu)
    }{
        \int dE_\nu d\Emu d\costheta \, \dsigma(E_\nu) \, \Phi(E_\nu)
    } ~ .
\end{equation}
We define the ND and FD \emph{true} densities $\pND$ and $\pFD$ by this expression evaluated with $\Phi_\mathrm{ND}$ and $\Phi_\mathrm{FD}$, respectively.
Evaluating Eq.~\eqref{eq:marginalize} with the modeled cross section in place of the true one defines the \emph{model} densities $\qND$ and $\qFD$. Note that our method works entirely with normalized densities $p(E_\ell, \cos\theta)$ at both the ND and FD, and so normalization information is completely discarded.
Our analysis is thus fully insensitive to the flux normalization uncertainty; flux shape uncertainties are discussed in Sec.~\ref{sec:flux-shape} below.

\section{Learning the cross section}
\label{sec:learning-the-xsec}
We construct and train a simple NN parameterization of the structure functions to provide a data-driven model of the cross section.
In particular, combining the known kinematic coefficients in Eq.~\eqref{eq:nuAd2-unique} with a NN parametrization of the three (combined) structure functions $W_i(x,q^2)$ gives an expressive model for $\dsigma$ which can be evaluated for arbitrary kinematics.
We train the model by tuning its parameters so that $\qND \approx \pND$ as closely as possible.
To focus on the more important issues of finite FD statistics and whether the cross section may be inferred in principle,
we first assume a perfect near detector and infinite ND statistics, i.e., we take $\pND$ to be known exactly with no noise.
Effects from finite ND statistics are subsequently included along with flux shape and energy resolution uncertainties in systematic uncertainty studies in Sec.~\ref{sec:nd-stats} below.

\begin{figure}[t]
  \centering
  \includegraphics[width=\linewidth]{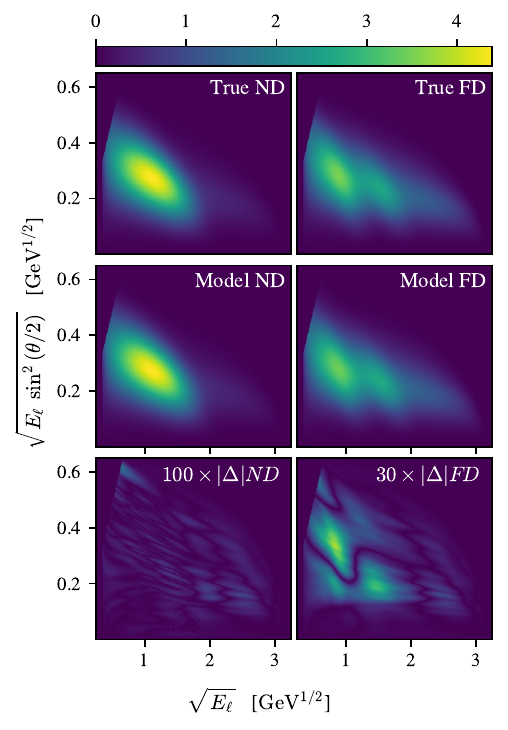}
  \caption{
    Event distributions at the ND (left) and FD (right) as predicted using either the true (top) or learned model (middle) cross sections and the true ND and FD fluxes.
    Differences between true and learned cross sections rescaled for visibility are also shown (bottom).   
    }
  
  \label{fig:marginals}
\end{figure}

The design of the training procedure is guided by the nontrivial physical requirements that the cross section be non-negative, but it decomposes into structure functions that may run negative.
These cannot be simultaneously satisfied by construction of the model, and must instead be enforced by training.
We therefore require a loss that is well-defined for negative values of $q_\mathrm{ND}$, which excludes common information-theoretic losses like the KL divergence~\cite{Kullback:1951}.
Instead, we use the mean squared error, 
\begin{equation}\label{eq:MSE_loss}
    \mathrm{MSE} = \displaystyle\int d\Emu d\costheta \Bigl[\pND(\Emu, \costheta) - \qND(\Emu, \costheta)\Bigr]^2.
\end{equation}
Because $\pND$ is non-negative, this choice drives $\qND$ to be non-negative without any additional regularization.

For computational expediency, we discretize all integrals on regular grids over $E_\nu$, $v_1 \equiv \sqrt{\Emu}$, and $v_2 \equiv \sqrt{\Emu \sin^2(\theta/2)}$.
Changing variables $(\Emu, \cos \theta) \rightarrow (v_1, v_2)$ gives more even distribution of the ND and FD densities, as visible in Fig.~\ref{fig:marginals}, and thus reduces discretization errors.
We note that for an at-scale application, there is no obstacle to the more principled approach of direct Monte Carlo integration over ND events, which, moreover, will obviate the need for any ND histogram construction.

This motivates our ML setup in the abstract.
Concretely, the results shown are for a model with the three $W_i$ parametrized as the three output channels of a single multi-layer perceptron (MLP) with two input channels for $x$ and $Q^2$, and 4 hidden layers of width 64 with LeakyReLU activations (see Fig.~\ref{fig:illustrate}).
For training, we use a $256 \times 128^2$ grid over $0 \leq E_\nu \leq 10~\mathrm{GeV}$, $0.25 \leq v_1/\mathrm{GeV}^{1/2} \leq 2.5$, and $0 \leq v_2/\mathrm{GeV}^{1/2} \leq 0.65$.
The integral defining the MSE loss is thus evaluated on a $128^2$ grid in $v_1$ and $v_2$.
We apply $10^4$ steps of the Adam optimizer~\cite{Kingma:2014vow} using default hyperparameters.  See Appendix~\ref{sec:ml-details} for more ML details.

The result is a close approximation of the true flux-averaged cross section, as apparent in the comparisons of Fig.~\ref{fig:marginals}.
The true and model ND densities are visually indistinguishable, with discrepancies at the $\sim$ 1\% level---as expected, given that this was the training objective.
Unlike in a real application, in the closure test we have access to the true $\Phi_\mathrm{FD}$ to compute and compare true and model densities, $p_\mathrm{FD}$ and $q_\mathrm{FD}$, at equivalently high resolution.
We see that the FD event density is nearly as well-modeled as the ND one, with discrepancies at the few-percent level, even though no FD information was provided during training.
The model thus enables good extrapolation from ND to FD kinematics.

\begin{figure*}
    \includegraphics[width=\textwidth]{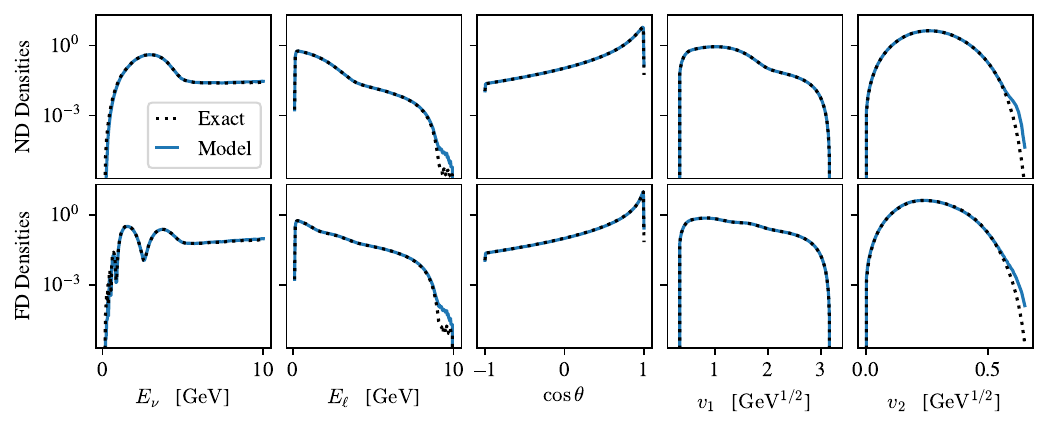}
    \caption{
    One-dimensional marginal event densities at the ND (top) and FD (bottom) as predicted using either the true (dashed black) or model (blue lines) cross sections and the true ND and FD fluxes.
    Computed from discretized integrals on a $256^3$ grids, specifically over $(E_\nu, \Emu, \costheta)$ for $\Emu$ and $\costheta$, and over $(E_\nu, v_1, v_2)$ for $v_1$ and $v_2$. 
    The marginals for $E_\nu$ can be computed equivalently using either grid.
    }
    \label{fig:marginal-comp-1d}
\end{figure*}

This same high-quality modeling at the ND and good extrapolation to the FD holds also in more detailed comparisons. For example, Fig.~\ref{fig:marginal-comp-1d} compares true versus model one-dimensional marginal densities of the various kinematic variables, $E_\nu$, $\Emu$, $\costheta$, $v_1$, and $v_2$.
Some mismodeling is visible in regions of lower event density, particularly in $\Emu$ and $v_2$.
However, the close agreement between true and model marginals in regions of high event density indicates excellent modeling of kinematically relevant parts of phase space.

\begin{figure*}[]
    \includegraphics[width=0.49\linewidth]{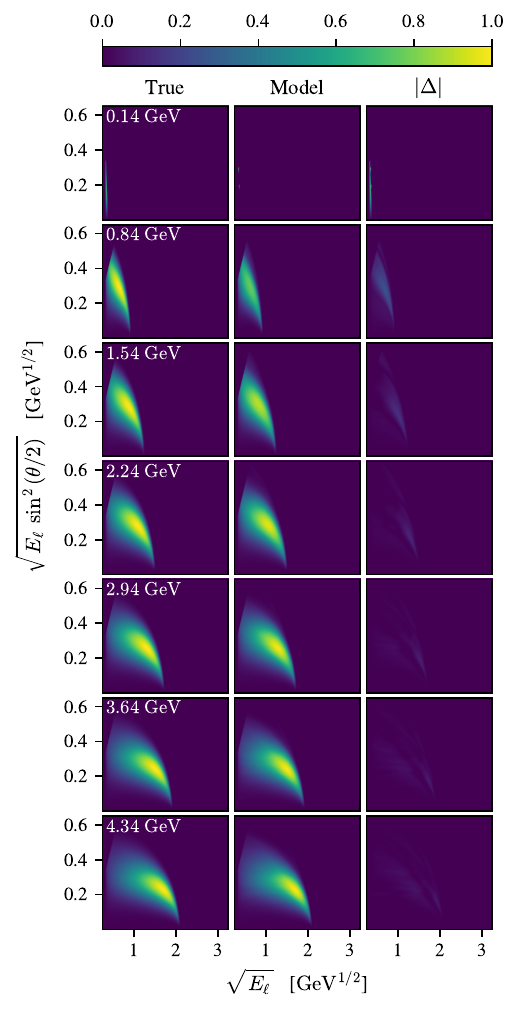}
    \includegraphics[width=0.49\linewidth]{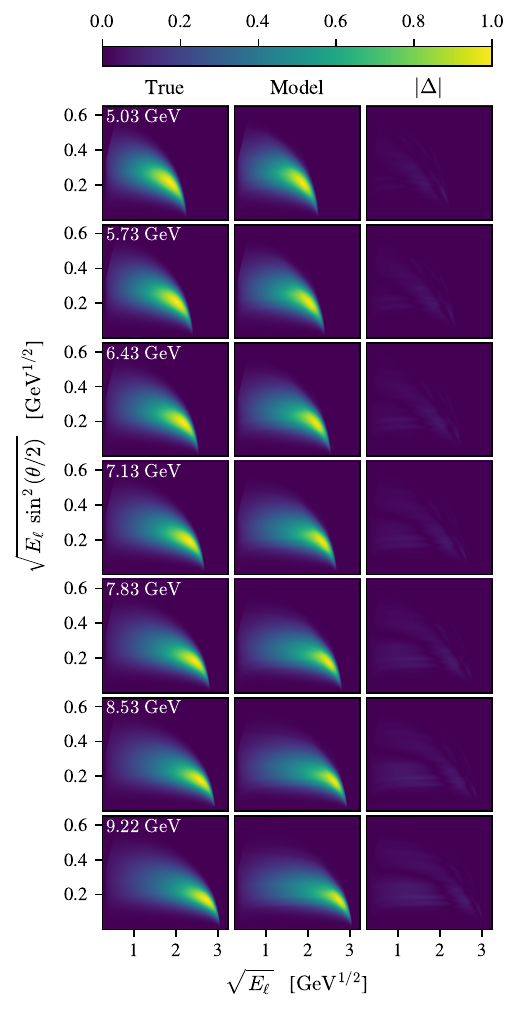}
    \caption{
        Comparisons of true and model cross sections along slices of fixed $E_\nu$ interpolating the full range of $0 \leq E_\nu \leq 10~\mathrm{GeV}$.
        Each cross section is first normalized as described in the text to remove an overall scale, then within each row the maximum value over either true or model is divided out of both.
        Not visible given this normalization convention is that $\dsigmav$ increases as a function of $E_\nu$, as shown in Fig.~\ref{fig:int-xsec-comp}.
    }
    \label{fig:xsec-comp}
\end{figure*}

The high quality of extrapolation to FD kinematics suggests good modeling of the underlying three-dimensional cross section. In the context of our closure test, where we have access to the true cross-section, we can verify this directly.
Figure~\ref{fig:xsec-comp} compares the true and model cross sections, evaluated on slices of fixed $E_\nu$ and shown for $(v_1, v_2)$ kinematics.
When rendered with the same colormap as the cross section, differences $|\Delta|$ are difficult to see at intermediate $E_\nu$. Small structured differences are apparent at low and high $E_\nu$. 

\begin{figure}
    \includegraphics[width=\linewidth]{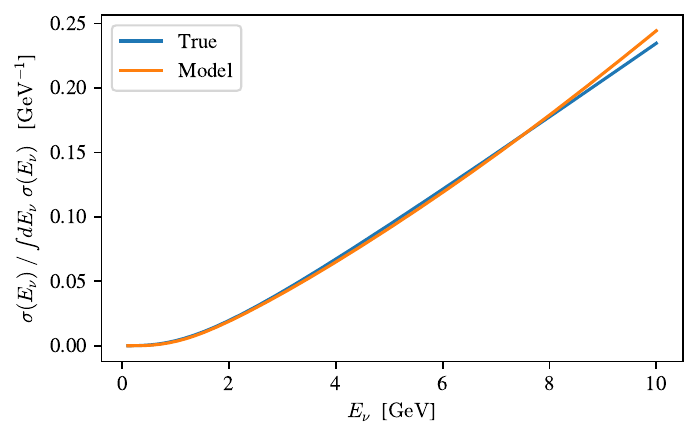}
    \caption{Comparison of true and model normalized (total) cross sections.}
    \label{fig:int-xsec-comp}
\end{figure}

Considering the total cross section $\sigma(E_\nu) = \int d\Emu d\costheta \, \dsigma(E_\nu)$ allows the size of these $E_\nu$-dependent discrepancies to be quantified.
To remove the overall scale ambiguity, we consider the normalized cross section $\sigma(E_\nu) / \int dE_\nu \sigma(E_\nu)$, where the integral is evaluated over the full kinematic range $0 \leq E_\nu \leq 10~\mathrm{GeV}$ of the toy model.
Note that this definition amounts to simply integrating over the slices shown in Fig.~\ref{fig:xsec-comp} and normalizing.
Figure~\ref{fig:int-xsec-comp} compares this quantity as computed using the true and model cross sections, confirming good agreement over most of the kinematic range, with deviations increasing at high $E_\nu$. 

Note that although the neural network models the cross section quite well, assessing the quality of modeling of the trained structure functions is more nuanced. 
See Appendix~\ref{sec:sf_comp} for a detailed analysis.

The various comparisons of this section indicate that high-quality modeling of the cross-section has been achieved, at least in the parts relevant for far-detector kinematics.
However, it remains to be demonstrated that this cross-section model is of sufficient quality to enable an oscillation analysis at the far detector.
This can be verified more directly by carrying out an oscillation analysis.


\section{Neutrino oscillation analysis}\label{sec:osc}

In an oscillation analysis, a model for the cross section is combined with a parametrization of the far-detector flux in terms of the unknown oscillation parameters, which are then constrained by comparison with the observed FD event distribution.
To benchmark the ideal case of perfect cross-section knowledge, we first consider the scenario for the far-detector analysis in which the true cross section is known exactly. 
This establishes a theoretical limit on the precision of the oscillation measurement when limited only by finite FD statistics.
In the practical scenario, we must instead rely on an imperfect model of the cross section constrained by ND data.
A successful ML model should therefore yield results for the oscillation analysis that are comparable to the ideal benchmark.

The flux of muon neutrinos reaching the far detector, $\Phi_\mathrm{FD}(E_\nu)$, can be modeled by $\widetilde{\Phi}_\mathrm{FD}(E_\nu; \omega) \equiv \Phi_{\rm ND}(E_{\nu}) P_{\mu\mu}(E_{\nu};\omega)$, where the muon neutrino survival probability $P_{\mu\mu}$ depends on the oscillation parameters collectively denoted $\omega$.
Because we consider only the muon disappearance channel, we fit solely for $\sin^2 (2\theta_{23})$ and $\Delta m_{31}^2$, with all other parameters fixed to their ``true'' values. 
We also restrict to the normal mass ordering in our analysis.
Incorporating the electron appearance channel will be essential in more comprehensive analyses to constrain the remaining oscillation parameters.

For the ideal case, the true cross section is combined with $\widetilde{\Phi}_\mathrm{FD}$ to define the FD event distribution, $\pFDtilde(\Emu, \costheta; \omega)$.
For the model case, the model cross section similarly defines an FD event distribution $\qFDtilde(\Emu, \costheta; \omega)$.
We then use maximum likelihood estimation (MLE) to infer the oscillation parameters $\omega$,
i.e.,~for a sample of $N$ far-detector events $\{\Emu^{(i)}, \costheta^{(i)}\}$ distributed per $\pFD$, we take $\mathrm{argmax}_\omega \, \mathcal{L}(\omega)$ where
\begin{equation}
\label{eq:FD-likelihood}
    \mathcal{L}(\omega) = \prod_{i=1}^N \pFDtilde(\Emu^{(i)}, \costheta^{(i)}; \omega),
\end{equation}
for the true cross section, and similarly for the model cross section with $\pFDtilde \rightarrow \qFDtilde$.
During FD inference, we define the model cross section with would-be negative values ($\sim 3\%$) clamped to zero.
We evaluate Eq.~\eqref{eq:FD-likelihood} over 6200 simulated events sampled from $\pFD$, matching the FD statistics expected after 3.5 years of running DUNE in neutrino mode~\cite{DUNE:2020ypp}.
We employ bootstrap resampling~\cite{Efron:1982,DiCiccio:1996,Davison:1997} to study uncertainty by generating $25000$ synthetic datasets, each by drawing 6200 samples with replacement from the original, and computing the MLE estimate in each.

\begin{figure}[t]
  \centering
  \includegraphics[width=\linewidth]{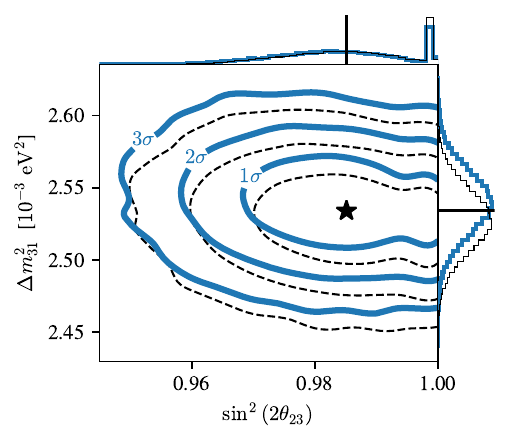}
  \caption{
    Confidence intervals inferred using either the true cross section (dashed black) or the learned model one (solid blue) with 6200 far-detector events, determined by bootstrapping $25000$ times through far-detector likelihood maximization.
    The stars and vertical lines indicate the true values.
    The histograms over bootstrap samples at the edges represent the marginal distributions of each inferred parameter.
    Smooth contour lines are computed from a kernel density estimate (KDE) constructed from the maximum-likelihood oscillation parameters $\omega$ computed for each bootstrap.
    \vspace{-10pt}
  }
  \label{fig:FD_inference}
\end{figure}

Figure~\ref{fig:FD_inference} shows confidence intervals (CIs) constructed from the resulting bootstrapped estimates of $\sin^2 (2\theta_{23})$ and $\Delta m^2_{31}$.
Confidence intervals computed using the true cross section and with the learned model
are of similar shape and extent, indicating good estimation of uncertainties with no artificial reduction due to mismodeling.
The tall bin at the right of the $\sin^2(2\theta_{23})$ histogram in Fig.~\ref{fig:FD_inference} can be attributed to the octant degeneracy. 

\begin{figure*}
    \includegraphics[width=\textwidth]{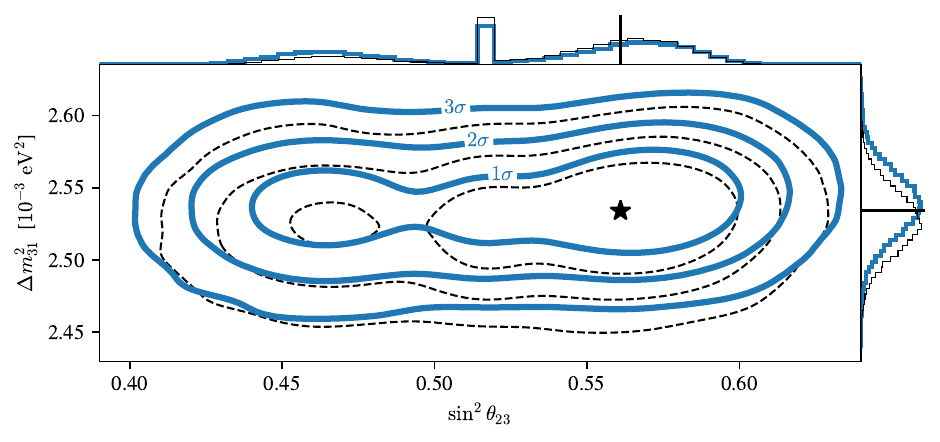}
    \caption{
    Confidence intervals as in Fig.~\ref{fig:FD_inference}, but in the $\sin^2\theta_{23}-\Delta m^2_{31}$ plane.
    Dashed black lines indicate results obtained using the true cross section and solid blue lines results obtained with the learned model one, determined by bootstrapping $25000$ times through far-detector likelihood maximization.
    The stars and vertical lines indicate the true values.
    The histograms over bootstrap samples at the edges represent the marginal distributions of each inferred parameter.
    Contour lines are computed using a kernel density estimate (KDE) over the bootstrap samples.
    }
    \label{fig:FD_inference_s23sq}
\end{figure*}

The octant degeneracy arises due to the two near-degenerate minima in $\sin^2 \theta_{23}$, which are difficult to resolve with only muon disappearance samples.
The variable $\sin^2(2 \theta_{23})$ is insensitive to the octant by construction and largely mitigates the complicates arising from this near-degeneracy.
For comparison, Fig.~\ref{fig:FD_inference_s23sq} presents the oscillation analysis for $\sin^2 \theta_{23}$ instead.
The confidence intervals show clear bimodality, with little preference for either mode.
The unusually tall bin in the marginal histogram in $\sin^2 \theta_{23}$ indicates that in a large fraction of bootstraps, the two minima are not resolved from one another (i.e., single-welled vs.~double-welled) such that MLE finds an intermediate value.
We note, however, that these issues arise even when using the true cross section and Figs.~\ref{fig:FD_inference} and \ref{fig:FD_inference_s23sq} demonstrate that the learned cross section accurately describes the same effects visible with the true cross section. 

\section{Systematic uncertainties}

With an idealized experimental setup, we have shown that our machine-learned cross-section model can, in an oscillation analysis, reach nearly the same precision as would be possible with perfect prior knowledge of the cross section. As a result, the uncertainty shown in Fig.~\ref{fig:FD_inference} is purely statistical and due to finite FD statistics. To understand the feasibility of this approach in realistic conditions, we must also assess the relevant systematic uncertainties. These broadly fall into two categories: those arising from experimental effects and those introduced by the machine-learning procedure itself.
In this section, we lay the groundwork to study these effects in our analysis framework and incorporate them into our uncertainty quantification. We study several key sources of systematic uncertainty in isolation, then integrate them into a combined framework to yield more realistic total uncertainties. Although preliminary, these results pave a path towards a principled approach to treat systematic uncertainties in our learned setup.
Additional studies of hyperparameter dependence are presented in Appendix~\ref{sec:hyperparams}, but are inessential to the present study.


\subsection{Finite ND statistics}
\label{sec:nd-stats}

Recent projections~\cite{DUNE:2021tad} estimate that in 3.5 years of running in neutrino mode, DUNE will observe $N_\mathrm{ND} \sim 30 \times 10^6$ $\nu_\mu$ events at the near detector. 
To derive the results in Secs.~\ref{sec:learning-the-xsec} and \ref{sec:osc}, we assume that the expected DUNE statistics are sufficiently large that we can work in the limit of infinite ND statistics.
Here, we verify that this assumption is reasonable.
Sec.~\ref{sec:uq-roadmap} below discusses how these effects are incorporated into our preliminary systematic uncertainty budget. 

Working in the infinite ND statistics limit corresponds to training the model to fit the exact two-dimensional event distribution, $p_\mathrm{ND}(v_1, v_2)$, as defined near Eq.~\eqref{eq:marginalize}.
In practice, this means we compute the integral Eq.~\eqref{eq:marginalize} by discretizing $\dsigmav(v_1,v_2,E_\nu)$ and $\Phi_\mathrm{ND}(E_\nu)$ on a $128^2 \times 256$ grid over $v_1,v_2,E$, respectively, then summing over the $E_\nu$ dimension.
The result is used as the training data without any noise added or other deformations.

Evaluated on a $128^2$ grid, $p_\mathrm{ND}(v_1, v_2)$ closely approximates the $N_\mathrm{ND} \rightarrow \infty$ limit of a density-normalized histogram constructed from ND events.
Testing for finite ND statistics effects can thus be accomplished by training the model on finite-statistics histograms.
Generating $N_\mathrm{ND}$ samples and then bootstrapping through histogram construction is feasible, but expensive; we simulate this procedure by drawing finite-statistics histograms $p_{\mathrm{ND},\text{finite}}$ from the appropriate multinomial distribution, i.e., we sample from
\begin{equation}
\begin{split}
    &\{ p_{\mathrm{ND},\text{finite}}(v_1^{(g)}, v_2^{(g)}, E_\nu^{(g)}) \}_g \\
    &= \frac{1}{N_\mathrm{ND}} \mathrm{Multinomial}[\{ N_\mathrm{ND} \, p_\mathrm{ND}(v_1^{(g)}, v_2^{(g)}, E_\nu^{(g)}) \}_g] ~ ,
    \end{split}
\end{equation}
where $g$ indexes the $128^2$ grid points.

\begin{figure*}
    \includegraphics[]{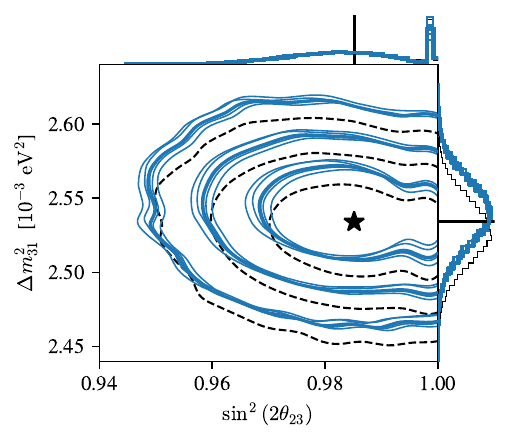}
    \includegraphics[]{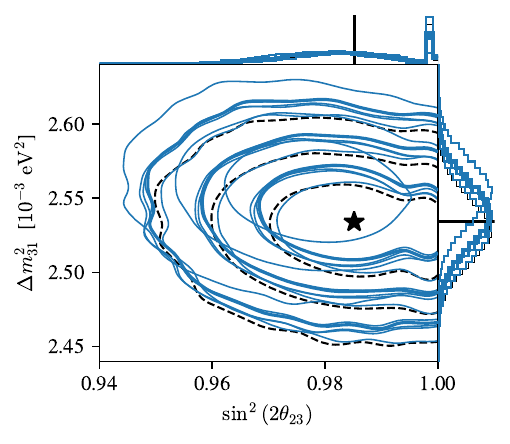}
    \caption{
    Confidence intervals as in Fig.~\ref{fig:FD_inference}, but including the effect of finite ND statistics at the projected level of $30 \times 10^6$ events~\cite{DUNE:2021tad} (left), and the effect of tenfold reduction in statistics to $3\times 10^6$ events (right).
    The dashed black line and thick blue band, obtained using the true cross section and the learned model, respectively, are reproduced from Fig.~\ref{fig:FD_inference}.
    The thin blue lines are computed using models trained on different simulated draws of ND events, with all other factors held fixed.
    }
    \label{fig:FD_inference_ND_stats}
\end{figure*}

We generate five such $p_{\mathrm{ND},\text{finite}}$ and train a model on each.
To isolate finite ND statistics effects, we keep all other factors fixed, including the model architecture and training hyperparameters as well as the initial model weights.
Visualizations comparing the resulting model event distributions and cross section are not noticeably distinct from those shown in Sec.~\ref{sec:learning-the-xsec}.
Instead, we compare the five finite-statistics models and the infinite-statistics at the level of the oscillation analysis.
We use the same set of 6200 FD events as used to produce Fig.~\ref{fig:FD_inference} with all models.

The result is the left plot of Fig.~\ref{fig:FD_inference_ND_stats}.
The confidence intervals derived agree well between the finite-ND-statistics models and the infinite-ND-statistics model.
While there is some fluctuation about the infinite-ND-statistics result, these variations are small relative to the extent of the confidence intervals and no systematic shift is apparent.
We thus conclude that finite ND statistics at the level of statistics expected for DUNE has only a minor effect in our analysis.

It is natural to ask whether the method remains applicable to lower-statistics preliminary datasets from DUNE or from other experiments.
To investigate, we repeat the same study with a factor 10 reduction of statistics, for a total of $3 \times 10^6$ ND events.
The right plot of Fig.~\ref{fig:FD_inference_ND_stats} shows the resulting effect on the oscillation analysis.
The qualitative success of the method persists.
However, it is clear that mismodeling due to finite ND statistics is a more significant effect at this reduced level of statistics, possibly on the same order as finite FD statistics.
We note that the size of this effect will necessarily depend on the fine details of our ML setup.
Future work will investigate whether improvements, including better choices of kinematic variables and methods for ND histogram construction, Monte Carlo evaluation of the loss function instead of histograms, and improved training dynamics, can partially mitigate these effects.


\subsection{Detector effects}
\label{sec:detector-effects}
\vspace{0.25em}

The results in Secs.~\ref{sec:learning-the-xsec} and \ref{sec:osc} were computed assuming perfect knowledge of each event $(\Emu,\costheta)$ is available.
However, in practice, no detector is perfect; finite resolution and other detector effects will distort each observation versus the true kinematics of the underlying event.
Here, we study the impact of such detector effects on our method.
These effects are also incorporated (at ND only) into our preliminary systematic uncertainty quantification scheme, as discussed in Sec.~\ref{sec:uq-roadmap} below.

Detector effects can be encoded by convolving the event distributions with a smearing kernel.
In full generality, this involves a kernel $S(\Emu, \costheta | \Emu', \costheta')$ which encodes the conditional density of the observed (i.e, smeared) event $(\Emu,\costheta)$ given the parameters of an underlying true event $(\Emu',\costheta')$.
In terms of this kernel, smeared event densities $p^S$ are obtained as
\begin{widetext}
\begin{equation}
    p^S(\Emu, \costheta) = \int d\Emu' \, d\costheta' ~ S(\Emu, \costheta | \Emu', \costheta') ~ p(\Emu', \costheta') ~ .
\end{equation}

DUNE is expected to reach $\sim$~4\% relative uncertainty on $E_\mu$~\cite{Grant2018DPFtalk,Marshalltalk}.
We encode it with the kernel
\begin{equation}
\begin{aligned}
    S(\Emu | \Emu') &= \frac{1}{\mathcal{N}_S(\Emu')} \exp \left[
        -\frac{1}{2 \sigma_{\Emu}^2} \left(\frac{\Emu-\Emu'}{\Emu'} \right)^2
    \right] \Theta(\Emu - \Emu^\mathrm{min}) \Theta(\Emu^\mathrm{max} - \Emu)
    ~,~~~\text{where}
    \\
    \mathcal{N}_S(\Emu') &= \sqrt{\frac{\pi}{2}} \Emu' ~ \sigma_{\Emu} \left[ 
        \mathrm{erf}\left( \frac{\Emu' - \Emu^\mathrm{min}}{\sqrt{2} \Emu' ~ \sigma_{\Emu}} \right)
        - \mathrm{erf}\left( \frac{\Emu' - \Emu^\mathrm{max}}{\sqrt{2} \Emu' ~ \sigma_{\Emu}} \right)
    \right] ~ ,
\end{aligned}
\end{equation}
\end{widetext}
with $\sigma_{\Emu} = 0.04$.
The Heaviside step functions $\Theta(x) = \{1 \text{ if } x \geq 0 \text{ else } 0 \}$ restrict $\Emu$ to the range defined in our toy model, $m_\ell \leq \Emu \leq 10~\mathrm{GeV}$.
The normalization factor $\mathcal{N}_S(\Emu')$ enforces $\int d\Emu ~ S(\Emu|\Emu') = 1$.
The estimated angular uncertainty of $1^\circ$ on $\theta$ is below the resolution of our discretized integrals, so we leave it for future investigation at larger computational scales where it may become relevant.

In practice, we evaluate integrals discretized over grids of 
$v_1 = \sqrt{\Emu}$ and $v_2 = \sqrt{\Emu \sin^2(\theta/2)}$, which requires translating $S(\Emu|\Emu')$ to $(v_1,v_2)$ kinematics.
Because $v_1$ is a function of $\Emu$ only, the appropriate construction is a function of $v_1$ only,
\begin{equation}
    S_v(v_1|v_1') = 2 v_1 S(v_1^2 | v_1^{\prime 2})
    ,
\end{equation}
where $2 v_1$ is the Jacobian factor from the change of variables.

Often, analyses of detector data aim to recover unsmeared events by ``unfolding'' the application of the smearing kernel on a per-event level.
This is not necessary in this framework used here, wherein detector events can be treated exclusively by ``forwards'' application of the kernel.
As described below, this is true for both cross-section inference at the near detector as well as the oscillation analysis at the far detector.

\begin{figure*}
    \includegraphics[]{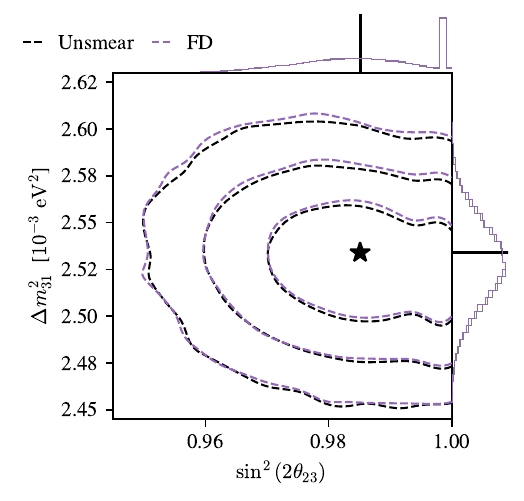}
    \includegraphics[]{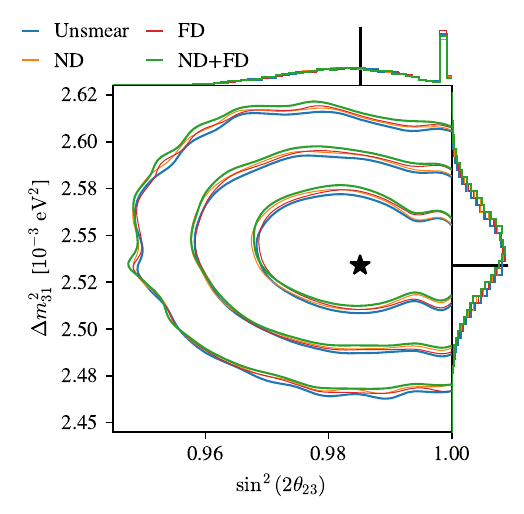}
    \caption{
    Confidence intervals (CIs) as in Fig.~\ref{fig:FD_inference}, but incorporating detector effects as described in the text.
    The left plot are CIs computed using the true cross section, with (FD) and without (Unsmear) detector smearing in the oscillation analysis at the far detector.
    The right plot are CIs computed with (FD, ND+FD) or without (Unsmear, ND) detector smearing at the far detector, using model cross sections learned either with (ND, ND+FD) or without (Unsmear, FD) detector smearing at the near detector.
    The dashed black line in the left plot and the blue line in the right plot (Unsmear) are reproduced from Fig.~\ref{fig:FD_inference}.
    }
    \label{fig:FD_inference_smear}
\end{figure*}

At the near detector, we apply $S(v_1|v_1')$ to the true event distribution $p(v_1, v_2)$ to obtain the true detector-smeared event distribution
\begin{equation}\label{eq:apply-ND-smear}
    p^S_\mathrm{ND}(v_1,v_2) = \int dv_1 ~ S(v_1|v_1') ~ p_\mathrm{ND}(v_1', v_2) ~ .
\end{equation}
This amounts to an additional discretized integral over $v_1$.
At each training step, we evaluate the model and construct a model event distribution $q_\mathrm{ND}(v_1',v_2)$.
We can then similarly apply $S_v(v_1|v_1')$ as in Eq.~\eqref{eq:apply-ND-smear} to obtain a smeared model density $q_\mathrm{ND}^S(v_1,v_2)$.
Training then minimizes the smeared MSE loss,
\begin{equation}
    \mathrm{MSE}^S = \int dv_1 dv_2 ~ \left[ p^S_\mathrm{ND}(v_1,v_2) - q^S_\mathrm{ND}(v_1,v_2) \right]^2 ~ .
\end{equation}
Visualizations comparing the resulting model event distributions and cross section are not noticeably distinct from those shown in Sec.~\ref{sec:learning-the-xsec}.

Just as at the near detector, detector smearing must be applied to both the data and model to carry out the oscillation analysis at the far detector.
For the data, we apply $S_v(v_1|v_1')$ to the 6200 events used in the oscillation analysis by adding to each a random offset sampled from a normal distribution with width $\sigma_{\Emu} \Emu$, rejecting and re-drawing any offset which would displace $\Emu$ outside the allowed range $[m_\ell, 10~\mathrm{GeV}]$.
To compute the likelihood, we must evaluate the smeared model densities $\tilde{p}_\mathrm{FD}^S$ and $\tilde{q}_\mathrm{FD}^S$ (defined with the true and learned cross-sections, respectively) for each event.
This introduces an additional integral, such that the total likelihood is
\begin{equation}
    \mathcal{L}(\omega) = \prod_{i=1}^N \int dv_1 ~ S_v(v_1^{(i)}|v_1') ~ \pFDtilde(v_1', v_2^{(i)}; \omega),
\end{equation}
for the true cross section, and similar for $\tilde{q}_\mathrm{FD}^S$ defined with the learned cross section.
We discretize the integral over $v_1'$ on a grid of 1024 points.

Figure~\ref{fig:FD_inference_smear} shows the effects of various combinations of detector smearing on the final oscillation analysis.
We find that detector effects are small relative to the statistical uncertainty due to finite far-detector statistics, whose scale is given by the spacing between confidence intervals.
If the true cross section is known exactly, then smearing need only be considered at the far detector; in the left plot, we see that FD smearing induces an insignificant shift towards larger $\Delta m^2_{31}$.
If the cross section must be learned, then smearing at the ND and FD can be considered separately.
The right plot assesses the four possible combinations.
FD smearing has a similar effect as with the true cross section.
The effects of ND smearing on the learned cross section induce a systematic shift similar in size and direction to FD smearing.
These effects compound, amounting to an $\approx O(0.1\sigma)$ effect in aggregate.


\subsection{Flux shape uncertainties}
\label{sec:flux-shape}

In our basic toy-model setup, we assume the ND flux $\Phi_\mathrm{ND}$ is known exactly with no systematic uncertainties.
In reality, our knowledge of the flux is imperfect.
Disagreement between the model flux at hand and the true flux underlying the data has effects on both oscillation analyses at the far detector as well as cross section inference at the near detector.
Here, we examine those consequences using a model of flux systematics based on estimates by the DUNE collaboration~\cite{DUNE:2021tad}.

Note that in a realistic setting, the flux and cross section may be simultaneously constrained using ND data.
It is necessary in any case to have two independent datasets in order to constrain both the flux and cross section; simultaneous fitting both the flux and cross section to a single distribution effectively leads to a black-box 3d  model of the cross section that will not accurately generalize to FD fluxes as discussed in Appendix~\ref{sec:3d_bad}.
The use of multiple exclusive hadronic channels to constrain both the flux and cross section is an  important avenue left for future work.
Here, we assume that the flux has been independently constrained using e.g.~neutrino-electron scattering data~\cite{Marshall:2019vdy} at a level of precision commensurate with DUNE collaboration flux uncertainty estimates discussed further below.

To treat flux uncertainties, we construct a ``flux systematics model'', a distribution over ND fluxes $\widetilde{\Phi}_\mathrm{ND}$ which encodes our uncertain knowledge of the true underlying flux $\Phi_\mathrm{ND}$.
In our treatment, the mock experimental data are held fixed while the fluxes used to construct model event densities are varied; this reflects experimental reality, where the data correspond to some unknown truth which must be modeled.
That is, mock data are always generated with respect to the true flux $\Phi_\mathrm{ND}$ of our toy model, while $q_\mathrm{ND}(\Emu,\costheta)$ and $\tilde{q}_\mathrm{FD}(\Emu,\costheta;\omega)$ are constructed using  $\widetilde{\Phi}_\mathrm{ND}$ sampled from the flux systematics model.
Examining the variation of results over different draws of $\widetilde{\Phi}_\mathrm{ND}$ allows flux shape uncertainties to be quantified.

\begin{figure}
    \includegraphics[width=\columnwidth]{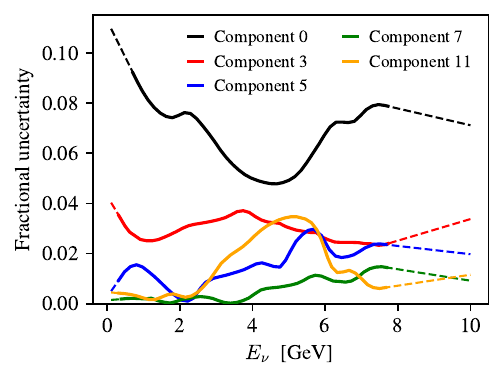}
    \caption{
        Fractional flux uncertainties used in our flux systematics model, taken from the principle components in Fig.~4 of Ref.~\cite{DUNE:2020jqi}; the dashed part of each line indicates linear extrapolations outside the range displayed in Ref.~\cite{DUNE:2020jqi}.
    }
    \label{fig:flux_fracs}
\end{figure}

\begin{figure*}
    \includegraphics[width=\linewidth]{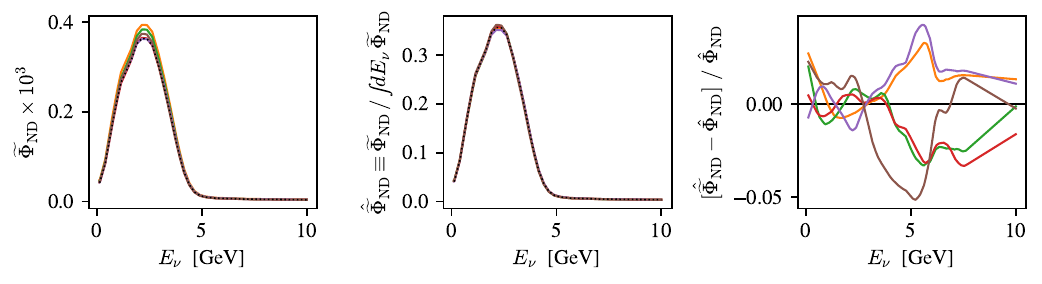}
    \caption{
        Comparisons of five draws $\widetilde{\Phi}_\mathrm{ND}$ from the flux systematics model (colored lines) with the true toy-model flux $\Phi_\mathrm{ND}$ (dashed black lines).
        The left panel shows each flux with its natural normalization.
        The middle panel shows unit-normalized flux densities.
        The right panel shows relative deviations of the five $\widetilde{\Phi}_\mathrm{ND}$ from the true $\Phi_\mathrm{ND}$; the comparison is made between unit-normalized fluxes to extract the part relevant to our normalization-insensitive analysis.
    }
    \label{fig:flux_samps}
\end{figure*}

To construct the flux systematics model, we use the fractional uncertainties from Fig.~4 of Ref.~\cite{DUNE:2020jqi}, specifically those constructed from five of the largest principal components of the flux covariance matrix studied there.
To cover the full range $[0,10]~\mathrm{GeV}$, we linearly extrapolate from the edges of the data shown.
The resulting fractional uncertainties $f_i(E_\nu)$ for $i \in {1,\ldots,5}$ are shown in Fig.~\ref{fig:flux_fracs}.
As discussed in Ref.~\cite{DUNE:2020jqi}, each of these curves are associated with independent nuisance parameters.
To sample from the systematics model, we therefore draw a set of 5 amplitudes $a_i$ from independent unit normal distributions, then construct a flux as
\begin{equation}
    \widetilde{\Phi}_\mathrm{ND}(E_\nu) = \Phi_\mathrm{ND}(E_\nu) \left[ 1 + \sum_{i=1}^5 a_i f_i(E_\nu) \right] ~ .
\end{equation}

The studies in this section use five draws from the resulting flux model, shown in Fig.~\ref{fig:flux_samps} in comparison with $\Phi_\mathrm{ND}$.
Our analysis is insensitive to the overall normalization of the flux, so the relevant comparison is between unit-normalized flux densities.
For typical draws, deviations from the true underlying flux are visible but less than $\sim 5\%$ as shown in the right panel of Fig.~\ref{fig:flux_samps}.
The reduction versus the typical scale in Fig.~\ref{fig:flux_fracs} is due to a combination of the removal of the overall normalization and because the average magnitude of draws from a unit normal distribution is less than one.

\begin{figure*}
    \includegraphics[]{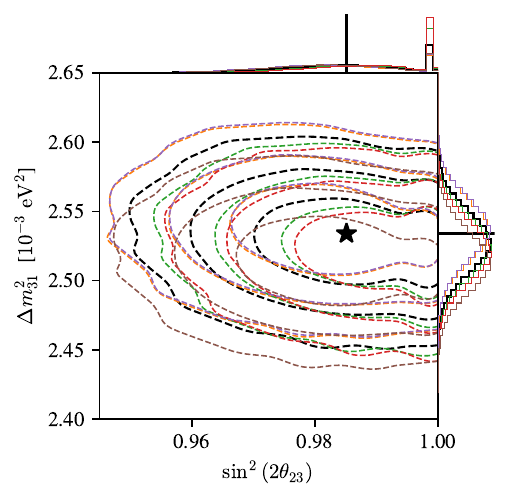}
    \includegraphics[]{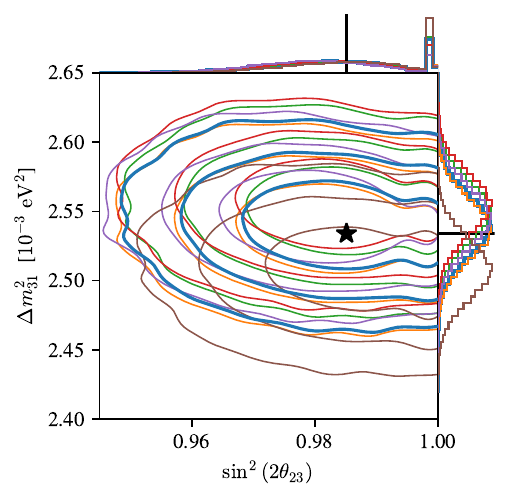}
    \caption{
        Confidence intervals (CIs) as in Fig.~\ref{fig:FD_inference}, but incorporating flux shape uncertainties as described in the text.
        The left plot are CIs computed using the true cross section of the toy model, for five different draws from the flux systematics model.
        For the same five draws, the right plot are CIs computed using model cross sections learned with mismodeled fluxes and with the same fluxes in the oscillation analysis.
        The dashed black lines (left panel) and thick blue lines (right panel) are results computed assuming perfect flux knowledge, reproduced from Fig.~\ref{fig:FD_inference}. 
    }
    \label{fig:flux_CIs}
\end{figure*}

To quantify the effect of flux shape uncertainties on the oscillation analysis, we keep the same fixed dataset generated using the true flux $\Phi_\mathrm{FD}$ as in Sec.~\ref{sec:osc}, but perform the likelihood maximizations using fluxes drawn from the model, i.e.,
\begin{equation}
    \widetilde{\Phi}_\mathrm{FD}(E_\nu; \omega) = \widetilde{\Phi}_\mathrm{ND}(E_\nu) \, P_{\mu\mu}(E_\nu; \omega) ~ ,
\end{equation}
defined with $\widetilde{\Phi}_\mathrm{ND}$ instead of the true $\Phi_\mathrm{ND}$.
To study the effect in isolation, we first consider oscillation analyses using the exact cross section from the toy model.
The left panel of Fig.~\ref{fig:flux_CIs} shows the results in comparison to those of Fig.~\ref{fig:FD_inference}.
The effect is clearly substantial, with the variation between the different sets of confidence intervals for each flux of roughly the same size as uncertainties from finite FD statistics.

Flux shape mismodeling will also affect the quality of the cross section model learned from ND data.
To reflect mismodeling of the underlying data, we use the true event density $p_\mathrm{ND}(\Emu,\costheta)$ generated with $\Phi_\mathrm{ND}$ as the training data, but use the model flux $\widetilde{\Phi}_\mathrm{ND}$ when constructing the learned model for the event density, i.e.,
\begin{equation}
    q_\mathrm{ND}(\Emu,\costheta)
    = \frac{ 
        \int dE_\nu \, \widetilde{\dsigma}(E_\nu) \, \widetilde{\Phi}_\mathrm{ND}(E_\nu)
    }{
        \int dE_\nu d\Emu d\costheta \, \widetilde{\dsigma}(E_\nu) \, \widetilde{\Phi}_\mathrm{ND}
    } ~ ,
\end{equation}
defined with $\widetilde{\Phi}_\mathrm{ND}$ instead of the true $\Phi_\mathrm{ND}$.
We train a model for each of the five flux draws, keeping all other factors fixed as in Sec.~\ref{sec:learning-the-xsec}.
Visualizations comparing the resulting model event distributions and cross section are not noticeably distinct from those shown in Sec.~\ref{sec:learning-the-xsec}.

We then perform an oscillation analysis with each of these five models.
In our toy model, we have assumed flat attenuation between the ND and FD, so we take the same flux $\widetilde{\Phi}_\mathrm{ND}$ for the FD oscillation analysis as used for ND training.
The right panel of Fig.~\ref{fig:flux_CIs} shows the resulting confidence intervals.
Comparing with the left panel, we see that the added effect of cross-section mismodeling results in noticeably larger variations in the confidence intervals.
However, it appears the dominant effect is in the FD inference problem, versus from mismodeling effects inherited from the ND inference problem.


\subsection{Model initialization systematics}\label{sec:initialization}

\begin{figure}
    \includegraphics[width=\columnwidth]{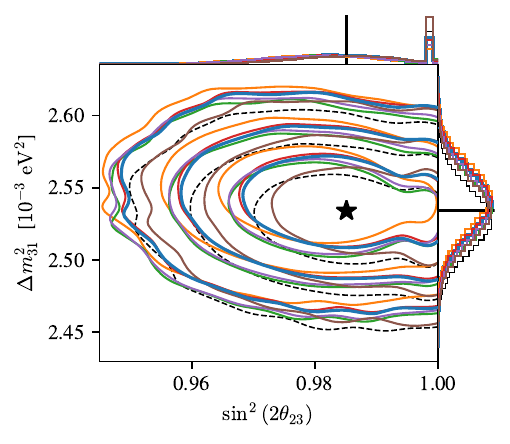}
    \caption{
    Confidence intervals as in Fig.~\ref{fig:FD_inference}, but for different random initializations of the model parameters (i.e.~different pseudorandom seeds).
    The dashed black line and thicker blue lines, obtained using the true cross section and the learned model, respectively, are reproduced from Fig.~\ref{fig:FD_inference}.
    The different-color lines are computed using models trained and evaluated identically, but initialized with different seeds.
    }
    \label{fig:FD_inference_seed}
\end{figure}

As implemented here, in the limit of infinite ND statistics and using discretized integrals, the only source of randomness in the inferred cross section is the random initialization of the model parameters.
Here, we examine the dependence of our results on different draws of these initial parameters---i.e., the variability with different choices of pseudorandom seed.

To investigate, we repeat identically the training procedure described above, but with five different pseudorandom seeds (for a total of 6 seeds sampled including the model featured in Sec.~\ref{sec:learning-the-xsec}).
All other factors are held fixed, including the model architecture and the discretization of the training data $p_\mathrm{ND}$.
In each case, we train for 10000 steps using the Adam optimizer and take the model with the best loss as the final output.
To get a sense for the variability in this selection, note that it takes the model after 9989, 9898, 9906, 9678, and 9895 steps on the five new seeds, compared with 9996 for the model featured in  Sec.~\ref{sec:learning-the-xsec}.
As discussed above, the overall normalization and thus the sign of the cross section is unconstrained, but may be identified straightforwardly post-hoc.
We find that it is $+1$ for 3 of the new models, and $-1$ for 2 of the new models; comparing with $-1$ for the initial model, it appears there is no strong bias towards one sign or another.
Visualizations comparing the resulting model event distributions and cross section are not noticeably distinct from those shown in Sec.~\ref{sec:learning-the-xsec}. 

Instead, we compare the different models at the level of the oscillation analysis in Fig.~\ref{fig:FD_inference_seed}.
The seed-dependent variation in the CIs is noticeably greater than the effects of finite ND statistics and detector smearing.
However, the qualitative conclusions above are unaffected by this variability.
Each set of model CIs are similar to the ones obtained with the true cross-section, and 
for all six models, the true values of $\sin^2(2\theta_{23})$ and $\Delta m^2_{31}$ lie well within the $1\sigma$ CI.
The same conclusions could be drawn from any fixed set of CIs.

It is natural to ask whether these observed seed-dependent variations can be somehow digested into an additional contribution to the uncertainty. Indeed, this procedure can provide some view on difficult-to-quantify modeling uncertainties.
To see this, note that these variations are of the same scale as the observed bias between confidence intervals computed using the learned cross section versus the true one, which can be attributed uniquely to mismodeling due to the simplicity of our setup.
This suggests that this procedure is giving some notion of the space of possible cross-sections consistent with our data.
As discussed in Sec.~\ref{sec:uq-roadmap} below, this variation can be incorporated into a systematic uncertainty estimate. Moreover, this provides a first step towards a Bayesian neural network treatment of uncertainties, as discussed below.


\subsection{Systematic uncertainty quantification}
\label{sec:uq-roadmap}

The results in Fig.~\ref{fig:FD_inference} demonstrate that it is possible to obtain closely comparable oscillation parameter estimates using a learned model cross section as if the true cross section were known exactly.
In particular, for the toy model used in the closure test, the $1\sigma$ confidence intervals obtained from the model contained the true oscillation parameters.
In this simplest possible sense, uncertainty quantification (UQ) has succeeded.
However, some deviations between true and model confidence intervals are apparent, and 
Fig.~\ref{fig:FD_inference} includes
no estimate of systematic uncertainties due to modeling or other effects.
This raises an important question: how can reliable UQ be guaranteed for this new class of methods, as necessary for applications to real-world neutrino experiments aiming to extract true parameters in nature?

First, it is useful to discuss what precisely is required of UQ.
In abstract, successful UQ can be defined in terms of calibration: if the true underlying value lies within the $1\sigma$ CI at least $68\%$ of the time, etc., then there is no danger of false confidence and uncertainty has been quantified appropriately.
It is acceptable, if undesirable, for UQ to be conservative: if the true value is within the $1\sigma$ CI 99\% of the time, there is no possibility of false confidence, but ideally a tighter error estimate could be obtained.
Defined in these terms, there is no unique correct or best uncertainty estimate for any method; any well-calibrated one is acceptable.

In this proof-of-principle study, we construct a basic systematic uncertainty quantification scheme and test its calibration in this sense. 
The combination of different stochastic effects on the ND inference problem---finite ND statistics, flux shape uncertainties, and seed dependence---induce a distribution of different cross section models.
The precise density of functions produced is a result of the interplay of this effect and the particular choice of architecture and training scheme.
Each model in the space of allowed cross section models will yield some predictions for oscillation parameters or, for finite FD statistics, some set of confidence intervals.
This space of predictions can be digested to produce an uncertainty estimate appropriately inflated to account for systematic effects.

To implement our UQ scheme, we proceed by training a sample of 25 models, incorporating all of the effects explored in the sensitivity studies above.
We assume each stochastic uncertainty is independent, motivating a simultaneous sampling approach.
That is, for each of the models, we draw a new:
\begin{itemize}[leftmargin=*]
\item Finite-statistics ND histogram $p_\mathrm{ND,finite}(\Emu,\costheta)$ for $30 \times 10^6$ samples, using the methods of Sec.~\ref{sec:nd-stats};
\item Model ND flux $\widetilde{\Phi}_\mathrm{ND}$ from the flux model of Sec.~\ref{sec:flux-shape};
\item Set of initial model parameters as in Sec.~\ref{sec:initialization}.
\end{itemize}
During training, we also account for detector smearing effects as in Sec.~\ref{sec:detector-effects}.
For computational expediency, we employ a twice-coarser grid for training than for the other studies, $128 \times 64 \times 64$ over $E_\nu$, $v_1$, and $v_2$; as found in Appendix~\ref{sec:hyperparams}, this coarsening has only a weak effect.
We keep the same training protocol as discussed above otherwise.

We employ a minor modification of the training scheme employed in Sec.~\ref{sec:learning-the-xsec} and detailed in Appendix~\ref{sec:ml-details}.
For $3/25 \approx 10\%$ of the draws, the models fail to converge for the first model initialization attempted.
In these cases, after 10k training steps, the loss remains substantially ($\approx 275\times$, $5\times$, $95\times$) larger than for a successful training run, and the model ND event density is visibly qualitatively different than the training data.
This sort of simple, easily diagnosed failures may arise from unlucky initializations which stall in a local minimum of the loss landscape under gradient descent.
Converged models can be obtained by running for longer, improving the training dynamics, or simply trying again with a different initialization.
We take this last approach for simplicity: whenever the loss fails to converge (diagnosable as e.g.~losses $>10\sigma$ higher than the typical converged value $\approx 0.097(7)$), we discard the present state, re-initialize the model, and train again.
We hold fixed all other factors, importantly including the draw from the systematics model, to ensure that the failure is associated with training dynamics and not with any systematic effect.
In all 3 cases of interest, a single restart yields a well-converged model within 10k steps.

With the procedure to sample models established, we must define some scheme for combining statistical and systematic errors.
One simple and conservative choice is simply to carry out an oscillation analysis to produce confidence intervals for each model and then take the union over all the resulting CIs.
However, in the limit of an infinite number of draws from the systematics model, the width of a union over all CIs may diverge.
Furthermore, such a union encodes only whether \emph{any} model from the distribution of possible models prefers those parameters, and not any information about how likely such a model is under the distribution induced by the systematics model.
A more nuanced construction is required to regulate this divergence and provide a meaningful uncertainty estimate.

To that end, we define the union over the $n$-$\sigma$ CIs as the set of all points included within at least a fraction
\begin{equation}
    f_n = \int_{n}^{\infty} dx \frac{1}{\sqrt{2\pi}} e^{-x^2/2} = \frac{1}{2}\left[ 1 + \text{erf}\left( \frac{n}{\sqrt{2}} \right) \right],
\end{equation}
of all the $n$-$\sigma$ CIs.
For $1\sigma$, this corresponds to all points within $\gtrsim 16\%$ of the $1\sigma$ CIs; for $2\sigma$, it corresponds to $\gtrsim 2.5\%$ of the $2\sigma$ CIs; etc.
Different choices of threshold are possible which would regulate the divergence.
This particular choice is motivated by the observation that it induces linear addition of statistical and systematic errors, i.e. $\sigma_\mathrm{stat.} + \sigma_\mathrm{syst.}$, $2\sigma_\mathrm{stat.} + 2\sigma_\mathrm{syst.}$, etc., in the limit where the width of the individual CIs are the same for all draws from the systematics model but their position fluctuates normally.

\begin{figure}[t]
  \centering
  \includegraphics[width=\linewidth]{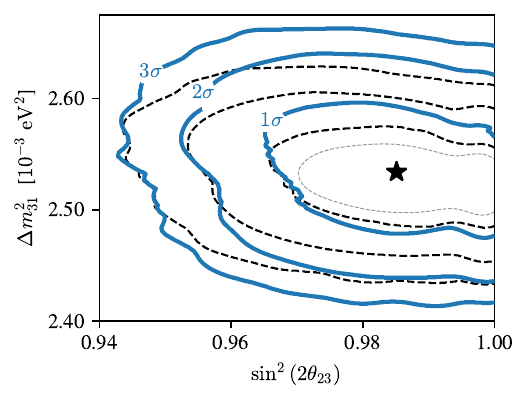}
  \caption{
    Analogous CIs to Fig.~\ref{fig:FD_inference} with true cross-section CIs (dashed black) including flux shape uncertainties and learned CIs including flux shape, finite ND statistics, and finite ND energy resolution systematics as well as NN initialization systematics (blue). The true cross section $1\sigma$ contour from Fig.~\ref{fig:FD_inference} is shown for comparison (gray dashed).
 \vspace{-10pt}    }
  \label{fig:FD_inference_uq}
\end{figure}

We apply this procedure to combine the results of oscillation analyses for the 25 draws of the systematic model and the learned cross sections that result from them.
As in Sec.~\ref{sec:flux-shape}, we use the same flux draw $\widetilde{\Phi}_\mathrm{ND}$ for the ND training and FD oscillation analysis.
For comparison, we also compute and combine CIs using the true cross section for the same draws, which is sensitive only to flux shape uncertainties.
We do not include detector smearing effects as in Sec.~\ref{sec:detector-effects} 
in the oscillation analyses due to its substantial computational expense; however, as seen in Sec.~\ref{sec:detector-effects}, this effect is subdominant and has similar effects for both true and model cross section.
With 25 draws, the union CIs include all points contained by the CIs of at least 4 models for $1\sigma$, and all points contained by any model's CIs for $2\sigma$ and $3\sigma$.
The resulting oscillation parameter constraints are shown in Fig.~\ref{fig:FD_inference_uq}.

Recalling the discussion of calibration above, we emphasize again that the results in Fig.~\ref{fig:FD_inference_uq} pass a basic but important check: the model CIs subsume or coincide with CIs computed with the true cross section, up to small violations which may be an effect of finite (25) draws from the systematics model.
This means that the modeling procedure produces oscillation parameter estimates that are at least as conservative as if cross section modeling were unnecessary.

More sophisticated UQ schemes are an active topic of research in AI/ML more broadly.
In particular, a roadmap for building on the basic systematic UQ scheme developed here is:
\begin{enumerate}[leftmargin=*]
    \item Explore the use of Bayesian neural network (BNN) constructions (see e.g.~\cite{arbel2024primer} for a review) to quantify the space of possible cross sections consistent with the data;
    \item With this space quantified, explore self-consistent calibration procedures in the spirit of Feldman-Cousins unfolding~\cite{Feldman:1997qc} and the plug-in principle from Efron's bootstrap~\cite{Efron:1982,DiCiccio:1996,Davison:1997} to estimate systematic uncertainties.
\end{enumerate}
Once a UQ procedure is defined, whether it is well-calibrated can be assessed using closure tests, ideally performed on as-close-to-physical examples as available.
Developing and testing a concrete UQ scheme along these lines will require conceptual and numerical work beyond the scope of this initial demonstration, anticipated to comprise a major component of future work.

\section{Discussion and conclusion}

We conclude that the method passes the closure test: using a model of the neutrino-nucleus cross section based on structure functions learned from data, oscillation parameters are inferred nearly as reliably as in the ideal benchmark case where the cross section is known {\it a priori}.

In more traditional analyses, significant bias in oscillation results can occur even after ND tuning in closure tests involving multiple event generators~\cite{DUNE:2021tad,Coyle:2025xjk}.
In comparison, in this closure test involving a simplified cross-section model, oscillation parameter CIs using a data-driven ML model subsume CIs using the true cross section without signs of bias.

Experimental systematic uncertainties---finite ND energy resolution\footnote{ Finite FD energy resolution can also affect oscillation studies but is subdominant in practice and has similar effects for the true and model cross-sections; see Sec.~\ref{sec:detector-effects}. It is not included in Fig.~\ref{fig:FD_inference_uq} for computational expediency.} and flux shape uncertainties---can impact ND training as well as FD oscillation parameter inference.
Additional systematic uncertainties arise from modeling uncertainties due to finite ND statistics and limited kinematic coverage, which in combination with the choice of training and architecture define a space of models consistent with the data.
We quantify these effects by taking appropriate unions of the CIs resulting from different random network initializations with experimental systematics included.
These ML and experimental systematics are included in learned cross-section oscillation results in 
Fig.~\ref{fig:FD_inference_uq} and compared to true cross-section results with flux shape uncertainties incorporated analogously.
Experimental systematics lead to $\approx 15\%$ and 75\% broadening of  $1\sigma$ CIs for $\sin^2(2\theta_{23})$ and $\Delta m^2_{31}$, respectively, while combined experimental and ML systematics leads to $\approx 20\%$ and 110\% broadening in the learned case.
This suggests that cross-section systematics can be similar in magnitude to experimental systematics in our approach.
More importantly, the model CIs approximately subsume the exact ones, suggesting that the modeling uncertainty estimate gives a conservative calibration.

Further development of data-driven methods can usefully complement generator-based approaches: having two independent methodologies with very different modeling uncertainties enables cross-validation of oscillation analyses.
Our results suggest several critical topics for future work.
Uncertainty quantification in ML approaches to solving inverse problems, of which learning cross sections is an example, is an active topic of research across the sciences~\cite{he2024surveyuncertaintyquantificationmethods} and more sophisticated methods should be explored.
Extending to more channels sensitive to different physics is also critical.
Exclusive final-state data will be necessary to fully exploit the unprecedented resolution of the DUNE experiment.
In particular, extensions of this work to semi-inclusive processes like electroweak pion production will be essential for leveraging the full power of DUNE datasets. These will require generalized structure function parameterizations involving low-energy versions of transverse-momentum-dependent parton distribution functions (TMDPDFs).
It is noteworthy that an inclusive analysis alone may already be sufficient for other experiments including T2K and Hyper-Kamiokande (although differences in ND and FD composition add other complications for these cases).

Incorporation of electron data is expected to resolve the octant degeneracy~\cite{DUNE:2020ypp}.
It will moreover resolve the degeneracies between the five structure functions in Eq.~\eqref{eq:nuAd2}, potentially allowing a better extraction of these quantities as physics targets in their own right.
The situation is more complicated for simultaneously analyzing neutrino and antineutrino data, which involve distinct structure functions for non-isoscalar nuclei such as argon; further data and/or theory inputs are required.
It may also be useful to incorporate data from multiple experiments with different kinematic coverage and physics priors from e.g.~perturbative QCD and nuclear effective field theories.
There are clear opportunities for synergy with the closely related NNSF$\nu$ approach~\cite{Candido:2023utz}, 
efforts to constrain NN models of response functions with electron scattering data~\cite{Sobczyk:2024ajv},
and experiments probing nuclear structure such as the Electron-Ion Collider (EIC)~\cite{Accardi:2012qut,AbdulKhalek:2021gbh,AbdulKhalek:2022hcn}.

DUNE and other accelerator neutrino experiments can provide a wealth of data enabling novel searches in the neutrino sector and new understanding of nonperturbative QCD in neutrino-nucleus scattering.
Data-driven cross-section modeling with ML enables accurate neutrino oscillation analyses without any of the nuclear theory assumptions entering standard, microscopic-theory-driven approaches.
Strong complementarity between data-driven and microscopic-theory-driven modeling will enable important cross checks on both approaches, e.g., tests for whether beyond-Standard-Model physics is being absorbed into data-driven cross-section models.
A combination of data-driven and microscopic-theory-driven approaches provides a promising route towards maximizing the discovery potential of DUNE.



\begin{acknowledgments}

We thank 
Minerba Betancourt, Arie Bodek, Steven Gardner, Alessandro Lovato, Pedro Machado, Luke Pickering, and Noemi Rocco for useful discussions.
This document was prepared using the resources of the Fermi National Accelerator Laboratory (Fermilab), a U.S. Department of Energy, Office of Science, Office of High Energy Physics HEP User Facility. Fermilab is managed by Fermi Forward Discovery Group, LLC, acting under Contract No. 89243024CSC000002.
The work of J.I. was supported by the U.S. Department of Energy, Office of Science, Office of Advanced Scientific Computing Research, Scientific Discovery through Advanced Computing (SciDAC-5) program, grant ``NeuCol''.
The work of K.T. is supported by DOE Grant KA2401045.
Numerical experiments and data analysis were performed using PyTorch~\cite{NEURIPS2019_9015}, NumPy~\cite{harris2020array}, SciPy~\cite{2020SciPy-NMeth}, pandas~\cite{jeff_reback_2020_3715232,mckinney-proc-scipy-2010}, gvar~\cite{peter_lepage_2020_4290884}, Mathematica~\cite{Mathematica}, and LHAPDF6~\cite{Buckley:2014ana}.
Figures were produced using matplotlib~\cite{Hunter:2007}.

\end{acknowledgments}

\appendix

\section{Failure of black-box 3d cross-section models}\label{sec:3d_bad}

\begin{figure}[t]
  \centering
  \includegraphics[width=\linewidth]{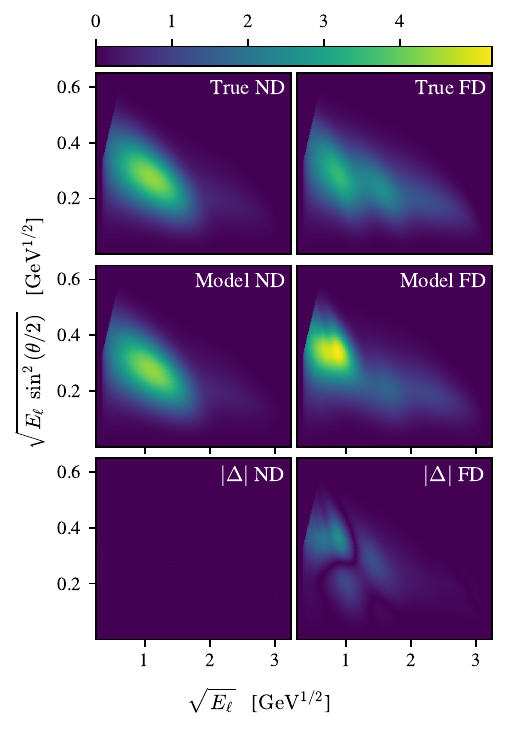}
  \caption{
    Failure of the black-box 3d modeling approach to extrapolate to FD kinematics when learned from the two-dimensional ND event distribution.
    Event distributions at the ND (top) and FD (middle) as predicted using either the true (left) or learned black-box 3d model (right) cross sections and the true ND and FD fluxes.
    The bottom row shows absolute differences between the event distributions on the same color scales.
  }
  \label{fig:marginals-3d}
\end{figure}

It is instructive to compare the performance of this SF-based model with a black-box three-dimensional (3d) model of the differential cross-section, including its $E_{\nu}$ dependence. 
In the black-box 3d model setup, we model the three-dimensional cross section ${\dsigmav}(E_\nu)$ directly.
Specifically, we parametrize a function which takes three inputs $(v_1, v_2, E_\nu)$ and returns a positive-definite value for the model $\widetilde{\dsigmav}(E_\nu)$ at those kinematics.
In particular, we employ an MLP architecture with three input channels for $(v_1, v_2, E_\nu)$, and 4 hidden layers of width 64 with LeakyReLU activations, and one output channel. Other than the number of input and output channels, this matches the architecture used for the structure function model.
We square the output of the MLP to guarantee a positive-definite cross section.
Other model architectures using for example normalizing flows can also be used to provide a positive-definite black-box 3d cross-section model; however, any black-box 3d cross-section will suffer from the same difficulties discussed below.

\begin{figure*}[t]
    \includegraphics[width=\linewidth]{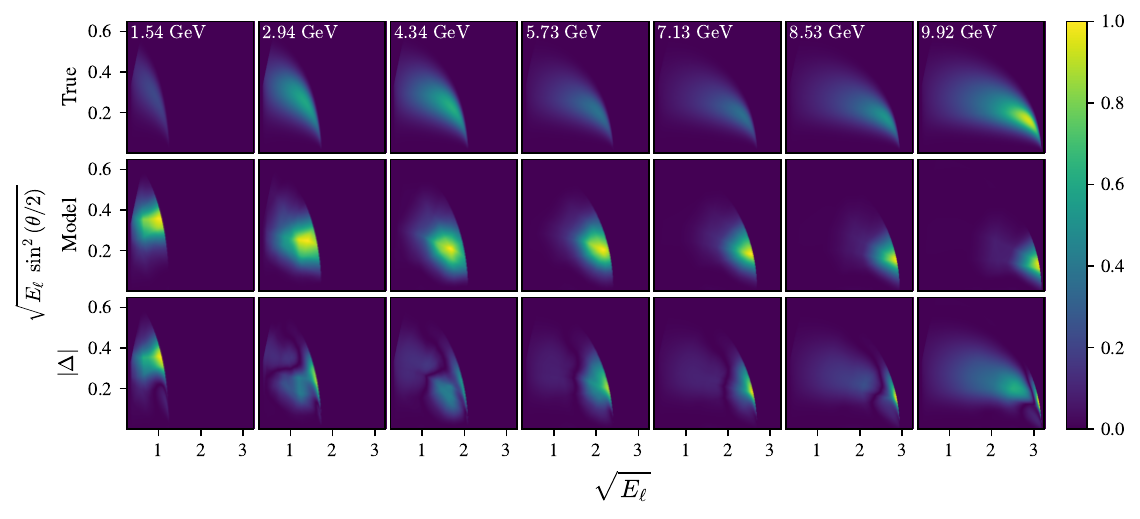}
    \caption{
        Failure of the black-box 3d modeling approach to learn the three-dimensional inclusive cross section. 
        Plots show comparisons of true (top) and black-box 3d model cross sections (middle) along slices of fixed $E_\nu$ interpolating the full range of $0 \leq E_\nu \leq 10~\mathrm{GeV}$.
        The bottom panel shows absolute differences between true and model on the same color scale.
        Each cross section is first normalized as described in the text to remove an overall scale, then within each row the maximum value over either true or model is divided out of both.
        Not visible given this normalization convention is that the true $\dsigmav$ increases as a function of $E_\nu$, as shown in Fig.~\ref{fig:int-xsec-comp}.
    }
    \label{fig:xsec-comp-3d}
\end{figure*}

We train exactly as with the structure function models.
In particular, we employ the MSE loss\footnote{Other options like the KL divergence are available in this case, in which the model is guaranteed to provide a positive-definite loss everywhere. We continue to use the MSE for a more direct comparison, but observe similar failures training with e.g.~the KL divergence.} and a $256 \times 128^2$ grid over $0 \leq E_\nu \leq 10~\mathrm{GeV}$, $0.25 \leq v_1/\mathrm{GeV}^{1/2} \leq 2.5$, and $0 \leq v_2/\mathrm{GeV}^{1/2} \leq 0.65$.
In $10^4$ steps of the Adam optimizer with default hyperparameters, the best MSE loss is achieved at step 6875.
Training dynamics are similar as with the structure function models, except with occasional larger instabilities in training that take many epochs to recover from.

Figure~\ref{fig:marginals-3d} shows the resulting model quality for ND and FD event distributions.
The black-box 3d model is learned directly on the two-dimensional ND event distribution. As can be seen, the quality of modeling at the ND is excellent: the true and model heatmaps are visually identical, and absolute discrepancies are not visible when rendered on the same colormap.

However, this cross-section model completely fails to extrapolate to FD kinematics.
Marginalizing against the true far-detector flux $\Phi_\mathrm{FD}$ to construct a model of the 2d FD event distribution, we obtain a FD event distribution which is clearly discrepant from the true one.
As shown in Fig.~\ref{fig:marginals-3d}, there are visible differences between the event distributions; the absolute deviation between true and model distributions is of a comparable scale to the densities themselves.
In Fig.~\ref{fig:xsec-comp-3d}, we interrogate this discrepancy more directly by examining slices of the full three-dimensional cross section.
Large, $O(1)$ differences are visible between true and model cross-sections across the full kinematic range.

We conclude that the learned cross section is entirely specific to the ND flux.
The only constraints imposed by training are that it reproduces the 2d ND event distribution when integrated against $\Phi_\mathrm{ND}$.
Without further constraints on the structure of the three-dimensional cross-section, there are flat directions in the loss which allow the three-dimensional structure of the model to vary arbitrarily.

This failure to learn an accurate representation of the underlying 3d cross-section can be easily understood analytically.
It is a consequence of the inference problem being irrecoverably \emph{ill-posed} when no constraints are made on the structure of the model.

The ill-posed nature of the 3d learning problem is straightforward to demonstrate analytically.
Given a target $d\sigma(\bm{v},h)$ that depends on ``visible'' parameters $\bm{v}$ and a ``hidden'' parameter $h$ (the neutrino energy above), a function of the hidden variable $\phi(h)$, and a known marginal distribution $P(\bm{v})$,
\begin{equation}\label{eq:3dlearning}
P(\bm{v}) = \frac{\int dh ~ d\sigma(\bm{v},h) \phi(h)}{\int d\bm{v} \, dh ~ d\sigma(\bm{v},h) \phi(h)} ~ ,
\end{equation}
the general version of the 3d learning problem is to model $\widetilde{d\sigma}(\bm{v},h) \approx d\sigma(\bm{v},h)$ only given knowledge of the marginal distribution $P(\bm{v})$.
It is possible to construct a family of exact outer-product solutions to the marginalized learning problem $\tilde{P}(\bm{v}) = P(\bm{v})$,
\begin{equation}
    \widetilde{d\sigma}(\bm{v},h) = \alpha P(\bm{v}) \phi(h),
\end{equation}
where $\alpha$ is any constant.
Inserting this model into Eq.~\eqref{eq:3dlearning} 
shows that $\tilde{P}(\bm{v}) = P(\bm{v})$ exactly,
\begin{align}
    \widetilde{P}(\bm{v}) 
    &= \frac{\int dh ~ \widetilde{d\sigma}(\bm{v},h) \phi(h)}
           {\int d\bm{v} dh ~ \widetilde{d\sigma}(\bm{v},h) \phi(h)} \\
    &= P(\bm{v}) \frac{\alpha \int dh \, \phi(h)^2}{ \alpha\int d\bm{v} \, P(\bm{v}) ~ \int dh \, \phi(h)^2}
    = P(\bm{v}) \nonumber
\end{align}
because $\int d\bm{v} \,P(\bm{v}) = 1$.
However, the outer-product is only one member of a (functionally) ambiguous family of exact solutions:
\begin{equation}\label{eq:3dgen}
    \widetilde{d\sigma}(\bm{v},h) = P(\bm{v}) f(h) + S_g(\bm{v},h)
\end{equation}
where $f(h)$ is \emph{any} one-dimensional function of $h$, and $S_g$ is any three-dimensional function satisfying
\begin{equation}
    \int dh \, S_h(\bm{v},h) \phi(h) = 0 ~ ,
\end{equation}
i.e., for which $\phi(h)$ is in the null space of $S_g$.
Such a function can be constructed from \emph{any} three-dimensional function $g(\bm{v},h)$ by simply projecting out the component along $\phi(h)$ as
\begin{equation}
    S_g(\bm{v},h) = g(\bm{v},h) - \frac{\int dh' \, g(\bm{v},h') \phi(h')}{\int dh' \, \phi(h')} . 
\end{equation}
We can plug in to check that this is an exact solution:
\begin{align}
    \widetilde{P}(\bm{v})
    &= \frac{
        P(\bm{v}) \int dh \, f(h) \phi(h)  \,+\, \int dh \, S_g(\bm{v},h) \phi(h)
    }{
        \int d\bm{v} \,P(\bm{v}) \int dh \, f(h) \phi(h)  \,+\, \int d\bm{v} \, dh \, S_g(\bm{v},h) \phi(h)
    } \nonumber \\
    &= P(\bm{v}) \frac{\int dh \, f(h) \phi(h)}{\int dh \, f(h) \phi(h)}
    = P(\bm{v}) 
\end{align}
using $\int d\bm{v} \,P(\bm{v}) = 1$ and $\int dh \, S_h(\bm{v},h) \phi(h) = 0$ in the first equality, and canceling the fractions in the second.
This still does not exhaustively cover the space of exact solutions; for instance arbitrary normalization constants could be further added to Eq.~\eqref{eq:3dgen}.
Considering unnormalized marginal rates, rather than normalized event densities, modifies the space but does not remove the existence of a nontrivial space or the associated functional ambiguities.

Without imposing further constraints on the 3d cross-section model $\widetilde{d\sigma}$, the existence of such (infinitely) ambiguous families of solutions means that any data-driven approach which seeks to optimize $\widetilde{P}(\bm{v}) \approx P(\bm{v})$ can only find, at best, some member of this exact family.
Any preference or apparent stability within this family can only be a consequence of an implicit regulator imposed by the choice of model architecture and fitting/training procedure, which serve to implicitly regulate this functional ambiguity.
For example, a particular neural network architecture and training procedure will prefer certain solutions to the problem; this preference carries no information about physics, and only information about the priors implicitly encoded by the ML.
Note that the dimension of the visible variables $\bm{v}$ does not affect the argument; thus, the same arguments apply for any exclusive and semi-inclusive process as well.
Incorporating additional physics constraints on the dependence of the cross-section on the marginalized variable $h$, e.g.~through the structure function parameterization explored elsewhere in this work, is required to provide a well-posed learning problem.

\section{Additional ML details}
\label{sec:ml-details}

In this appendix, we provide further details on the training procedure employed to produce the results in the main text.
For the model in Sec.~\ref{sec:learning-the-xsec}, training for 10000 steps takes approximately 14 minutes on an NVIDIA A100 GPU on Google Colab.

We note that there is no stochasticity in the training process.
Once the initial model weights are drawn randomly, training is fully deterministic.
This is because stochastic gradient descent is only stochastic when the loss (or more precisely, the gradients of the loss) are estimated stochastically.
This is not the case in the method explored in this work: the integrals defining the MSE loss are computed by discretizing them on a grid, rather than using a Monte Carlo estimator or random minibatching (i.e., taking random subsets of a finite training data set).

\begin{figure*}
    \includegraphics[width=\linewidth]{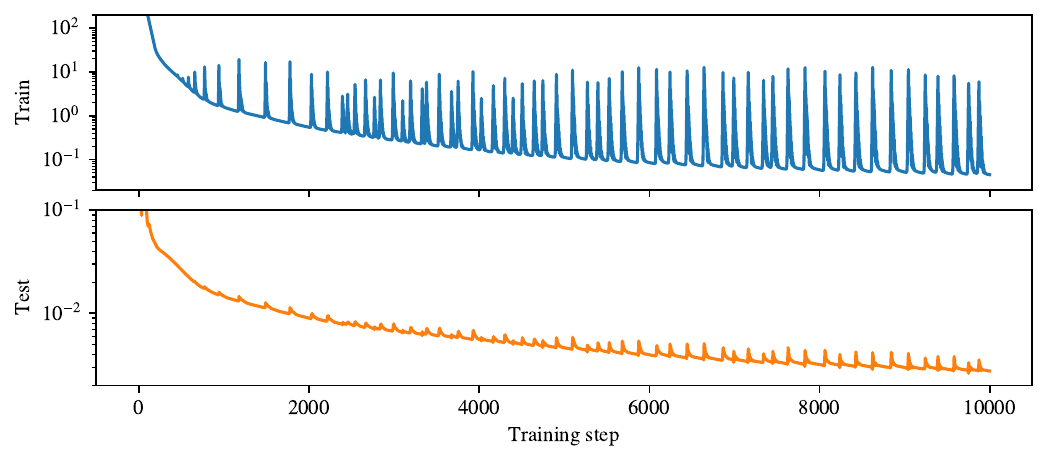}
    \caption{
        History of training loss (MSE, as defined in Eq.~\eqref{eq:MSE_loss}) and test loss (as discussed in the text) over the full course of model training.
        Note that large values at early training times are outside the range of the plot.
    }
    \label{fig:loss}
\end{figure*}

Figure~\ref{fig:loss} shows training and test loss curves.
The training loss is the MSE as defined in Eq.~\eqref{eq:MSE_loss}; the test loss is defined below.
The structure in the training loss curve---smooth descent interrupted by large spikes, then a decay back to the previous value---reflects an instability in the training process.
In the authors' experience, such instabilities often arise when training neural networks using non-stochastic losses.
This instability is not a practical problem.
Considering the lower envelope of the loss curve, it is clear that the quality of optimization continues to increase over time on the whole, with only transient disruptions.
To avoid finding a bad model if training concludes in the midst of such an event, we retain a copy of the model for the best loss observed thus far, and take that as the final output of training.
For the model of Sec.~\ref{sec:learning-the-xsec}, this occurs on the 9996th training step out of 10000.

Note that the ND and FD inference problems are each defined in terms of marginal distributions, i.e.~$\pND$, $\qND$, $\pFDtilde$, and $\qFDtilde$.
Computing a properly normalized marginal from a cross section and flux per Eq.~\eqref{eq:marginalize} requires divison by $\int dE_\nu d\Emu \costheta \, \dsigma \phi$.
This means that the marginal distributions are each invariant under overall rescalings of the flux or cross section.
Consequently, none of the inference problems considered here---neither learning the cross section at the ND nor the oscillation analysis at the FD---are sensitive to the overall scale of the flux or cross section.
Thus, the model cross section and structure functions can only be expected to be correct up to an overall scale factor, even in the limit of perfect modeling.

This overall scale factor may be negative. That this can occur does not pose any practical issue, because it can always be identified by examining the model cross section, which should be positive everywhere. 
In fact, the final model as initially trained is off by an overall sign, parameterizing a cross section which is negative (almost) everywhere.  
With no loss of rigor, we redefine the model after training as the outputs of the original function multiplied by $-1$. 
We emphasize that the inference problems are insensitive to this sign regardless, but it will be important if structure functions are a desired output.

While not possible when modeling an unknown cross section, in the toy-model setting, we know the true three-dimensional cross section and thus are able to compare it to the model one.
The test loss shown in Fig.~\ref{fig:loss} encodes this comparison.
Because the cross section can be learned only up to an overall scale, this comparison requires first defining normalized quantities.
In particular, we compute
\begin{equation}
    S_p(E_\nu, \Emu, \costheta) \equiv \frac{
        \dsigma(E_\nu) 
    }{
        \int dE_\nu d\Emu d\costheta ~ \dsigma(E_\nu)
    } ~ ,
\end{equation}
and similarly $S_q$ from the model cross section, from which the test loss is defined as
\begin{equation}\label{eq:test-loss}
    \int dE_\nu d\Emu \, d\costheta \, \left| S_p - S_q \right|^2 ~ .
\end{equation}
Note that these are written in terms of $\Emu, \costheta$ to avoid confusion, but in practice, we compute these integrals discretized over $v_1 = \sqrt{\Emu}$ and $v_2 = \sqrt{\Emu \sin^2(\theta/2)}$ kinematics.

The behavior of the test loss in Fig.~\ref{fig:loss} implies that training smoothly produces an increasingly high-quality approximation of the cross section across its full kinematic range in all three dimensions.
This is despite the fact that training only has access to $\pND$, a two-dimensional marginalization of the full three-dimensional object.
Interestingly, while some sign of the same instabilities observed in the train loss are visible in the test loss, the overall size of the effect is much reduced.
It may be interesting to determine the dynamics underlying this difference.



\section{Structure functions}
\label{sec:sf_comp}

\begin{figure*}
    \includegraphics[width=0.49\textwidth]{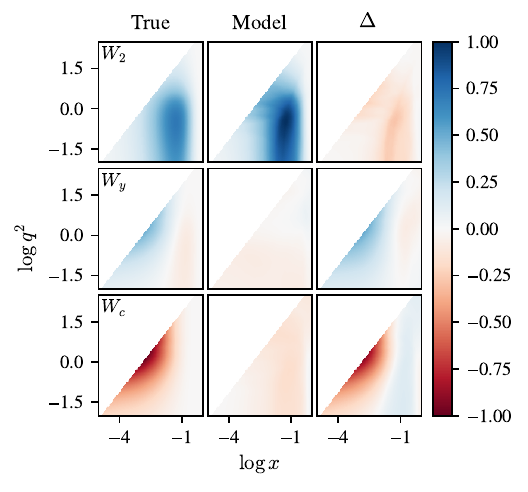}
    \includegraphics[width=0.49\textwidth]{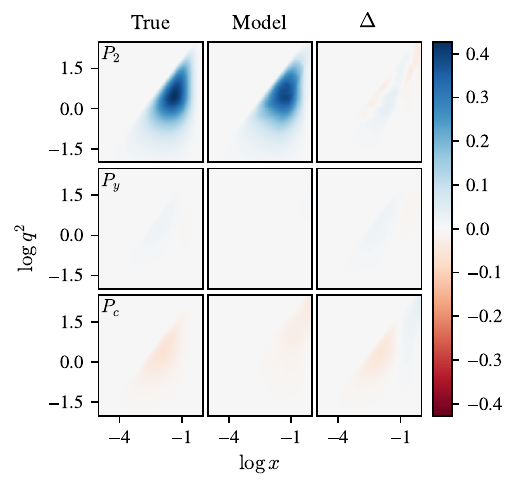}
    \caption{
        Comparison of true and model structure functions without (left) and with (right) weights accounting for which kinematic regions are well-constrained by the available ND data.
        The masked regions are fully unconstrained due to $E_\nu \leq 10~\mathrm{GeV}$.
        For each set of true and model structure functions separately, the overall scale of the $W_i$ is set by dividing the overall maximum.
        The relative scales between different structure functions are thus left intact and encoded in the colormaps. 
    }
    \label{fig:sf-comp}
\end{figure*}

As demonstrated, the cross section can be learned accurately over relevant kinematic ranges using ND data.
Ideally, the nuclear structure functions $W_i$ would also be a well-estimated physics output of the analysis.
In practice, however, they are not as obviously well-modeled as the cross section, as apparent from the left panel of Fig.~\ref{fig:sf-comp}.
Furthermore, the unclear relation between true and model $W_i$ naively seems inconsistent with the high quality of approximation of the cross section.

The source of this apparent discrepancy is that the ND marginal $\pND$ is related to the structure functions with nontrivially $(x,Q^2)$-dependent weights by the combination of the ND flux $\Phi_\mathrm{ND}$ and the kinematic factors of Eq.~\eqref{eq:nuAd2-unique}.
Via these weights, the ND data constrain only a small range of all $(x,Q^2)$, outside of which the model is free to vary without significantly affecting $q_\mathrm{ND} \approx p_\mathrm{ND}$ (and, critically, $q_\mathrm{FD} \approx p_\mathrm{FD}$).
For example, the whited-out regions in Fig.~\ref{fig:sf-comp} are those for which there are no constraints at all, due to the maximum $E_\nu=10~\mathrm{GeV}$ defined for the toy model.

Accounting for this kinematic weighting paints a clearer picture.
Because $\pND(\Emu, \costheta)$ is obtained by marginalizing over $E_\nu$, its relation to $W_i(x,Q^2)$ is nontrivial: each point in $(\Emu, \costheta)$ in principle constrains $W_i$ across the entire $(x, Q^2)$ domain.
While it may be interesting to explore applications of the four-dimensional weight function that this defines, a simpler option is available in the toy model setting: we may instead consider the three-dimensional ND event distribution $\mathcal{P}_\mathrm{ND}$ in $(x,y,Q^2)$ kinematics, and marginalize over $y$.
First, note that $\mathcal{P}_\mathrm{ND}$ may be decomposed into a contribution from each structure function:
\begin{widetext}
\begin{equation}
\begin{aligned}
    \mathcal{P}_\mathrm{ND}(x,y,Q^2) 
    \equiv \frac{1}{\mathcal{N}} \Phi_\mathrm{ND}(E_\nu(x,y,Q^2)) ~ \frac{d^2\sigma}{dx dy}(Q^2)
    &= \frac{1}{\mathcal{N}} \Phi_\mathrm{ND}(E_\nu(x,y,Q^2)) ~ \sum_i K_i(x,y,Q^2) W_i(x, Q^2)
    \\
    &= \sum_i \left[ \frac{1}{\mathcal{N}} \Phi_\mathrm{ND}(E_\nu(x,y,Q^2)) ~ K_i(x,y,Q^2) \right] W_i(x,Q^2)    
\end{aligned}
\end{equation}
\end{widetext}
where $K_i$ are the kinematic coefficients of the structure functions from Eq.~\eqref{eq:nuAd2-unique} and 
\begin{equation}
    \mathcal{N} \equiv \int dx dy dQ^2 ~ \frac{d^2\sigma}{dx dy}(Q^2) ~\Phi_\mathrm{ND}(E_\nu(x,y,Q^2)).
\end{equation}
Because $P_\mathrm{ND}$ is already normalized, marginalization over $y$ may be accomplished simply by integration, which allows further defining
\begin{equation}
\begin{split}
    \int dy ~ \mathcal{P}_\mathrm{ND}(x,y,Q^2) 
    &\equiv \sum_i \mathcal{K}_i(x,Q^2) W_i(x,Q^2) \\
    &\equiv \sum_i P_i(x,Q^2) ~ .
    \end{split}
\end{equation}
The $y$-marginalized coefficient functions
\begin{equation}
    \mathcal{K}_i(x,Q^2) \equiv \int dy \frac{1}{\mathcal{N}} \Phi_\mathrm{ND}(E_\nu(x,y,Q^2)) ~ K_i(x,y,Q^2)
\end{equation}
define $(x,Q^2)$-dependent weights which encode exactly which regions of the $W_i$ are relevant to ND kinematics.
Multiplying them on to $W_i$ defines $P_i(x,Q^2)$, which are the contributions associated with each $W_i$ to the total marginal $P(x,Q^2) \equiv \int dy ~ \mathcal{P}(x,y,Q^2)$, such that $\sum_i P_i = P$.

The right panel of Fig.~\ref{fig:sf-comp} compares the true and model structure functions with these kinematic weights applied to obtain $P_i(x,Q^2)$.
It is clear that $W_2$ is the overwhelming contribution, with $W_c$ and $W_y$ heavily kinematically suppressed.
This furthermore makes apparent that the kinematically relevant part of $W_2$ is well-modeled, explaining the high-quality approximation of the cross section.
More substantial mismodeling of $W_y$ and $W_c$ is faintly visible, but the overall scale of these effects are clearly subleading.

This analysis indicates that further refinements will be required if the structure functions themselves are the objects of interest, except for $W_2$ in a particular kinematic region.
While it may be possible to improve the extraction with additional methods developments, incorporating additional physics information provides a clear path to improvement.
For example, adding electron information allows in principle separately constraining all five $W_i$ of Eq.~\eqref{eq:nuAd2}. Furthermore, an approach similar to that of NNSF$\nu$~\cite{Candido:2023utz}, which fits SFs to multiple experiments with different systematic effects, would enable stronger constraints on different kinematical regions. 
However, many experiments use different targets; incorporating these data together requires some modeling of the dependence of the SFs on the proton and neutron number, and thus additional nuclear theory inputs. 


\section{Hyperparameter dependence}
\label{sec:hyperparams}

\begin{figure*}
    \includegraphics[width=\textwidth]{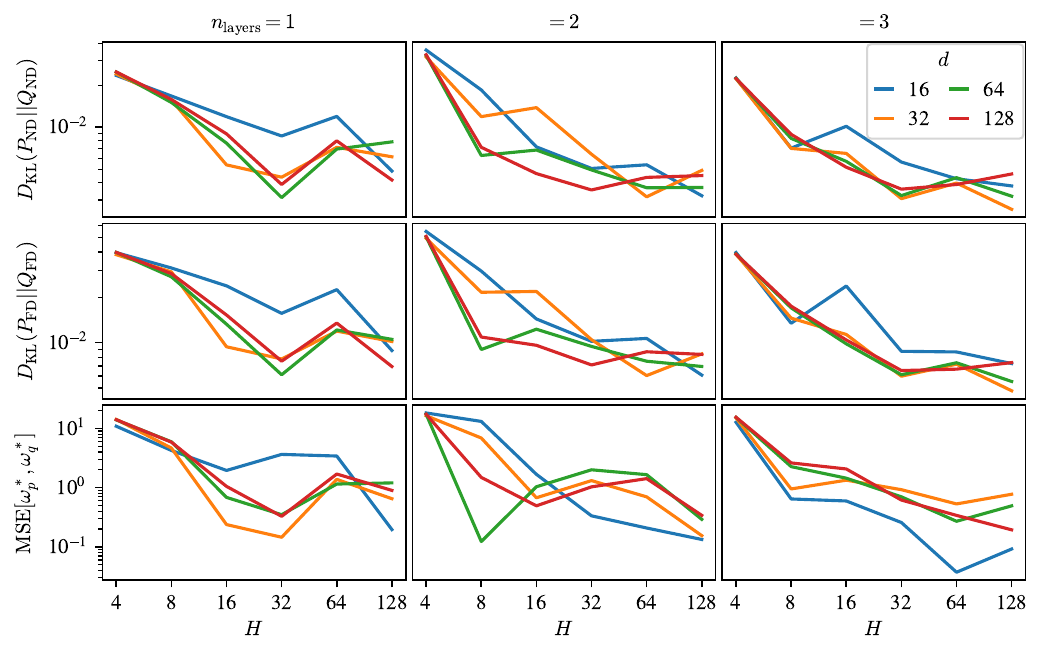}
    \caption{
    Scaling of different measures of model quality with hyperparameters.
    During training, all integrals are discretized on $d \times d \times 2d$ grids over $(v_1,v_2,E_\nu)$ (different color curves).
    During evaluation, a common finer discretization is used.
    Depth $n_\mathrm{layers}$ (columns) and hidden layer width $H$ define the size of the MLP used to parametrize the structure functions.
    As defined more precisely in the text, the top two rows show KL divergences, while the bottom row quantifies the deviation in maximum-likelihood parameter estimates using the model from those using the true toy-model cross section.
    }
    \label{fig:hyperparam_scaling}
\end{figure*}

It is natural to ask whether the results presented here depend on the particularities of the ML setup, i.e., the choice of architecture, training procedure, and all the finer-grained hyperparameters that define the precise procedure employed.
While an exhaustive exploration is intractable, it is straightforward to study dependence on select hyperparameters.
We perform a scan over three different hyperparameters of particular interest.
Two are architectural and define the size of the MLP used to parametrize the cross-section: the depth $n_\mathrm{layers}$ (more precisely, number of hidden layers) and the width $H$ (of the hidden layers).
The third, $d$, defines the resolution of the $d \times d \times 2d$ grid on $(v_1,v_2,E_\nu)$ over which discretized integrals are evaluated during training.

Figure~\ref{fig:hyperparam_scaling} shows different metrics of performance when scanning over $d \in \{ 16, 32, 64, 128 \}$, $n_\mathrm{layers} \in \{ 1, 2, 3\}$, and $H \in \{ 4, 8, 16, 32, 64, 128\}$.
The model featured in Fig.~\ref{fig:FD_inference} corresponds to $d = 128$, $n_\mathrm{layers} = 3$, and $H = 64$.
We show three different metrics, each of which presents a different comparison of the learned and true cross sections.

The first two metrics are KL divergences comparing the true and model three-dimensional event densities, denoted with uppercase $P$ and $Q$ to distinguish them from the two-dimensional event densities, at the near and far detectors:
\begin{equation}
    \begin{split}
    D_\mathrm{KL}(P_\mathrm{ND}||Q_\mathrm{ND}) =& \int dv_1 dv_2 dE_\nu ~ P_\mathrm{ND}(v_1,v_2,E_\nu) \\
    &\times\log \frac{P_\mathrm{ND}(v_1,v_2,E_\nu)}{Q_\mathrm{ND}(v_1,v_2,E_\nu)} ~ 
    \end{split}
\end{equation}
and similar for the FD.
$D_\mathrm{KL}(P||Q) = 0$ when $P = Q$ and greater otherwise.
The integrals are discretized on a $256^2 \times 512$ grid.
We have also examined and find similar pictures for MSEs of the same densities, as well as the test loss Eq.~\eqref{eq:test-loss}.

The third metric in Fig.~\ref{fig:hyperparam_scaling} quantifies performance in the oscillation analysis.
Specifically, it is an MSE of normalized distances between maximum-likelihood oscillation parameters obtained using the model ($q$ subscripts) and true ($p$ subscripts) cross section.
Evaluated over $B=5000$ bootstrap draws $b$ of the 6200 FD samples, it is
\begin{widetext}
\begin{equation}
    \mathrm{MSE}[\omega^*_p, \omega^*_q] \equiv \frac{1}{B} \sum_b \left[
        \left(\frac{
            [\sin^2(2\theta_{23})]^{(b)}_p - [\sin^2(2\theta_{23})]^{(b)}_q 
        }{
            \mathrm{Std}_{b'}\left[ [\sin^2(2\theta_{23})]^{(b')}_p \right]
        }\right)^2  + \left(\frac{
            [\Delta m_{31}^2]^{(b)}_p - [\Delta m_{31}^2]^{(b)}_q 
        }{
            \mathrm{Std}_{b'}\left[ [\Delta m_{31}^2]^{(b')}_p \right]
        }\right)^2
    \right] ~ ,
\end{equation}
\end{widetext}
where $\mathrm{Std}_{b'}$ indicates the standard deviation over bootstraps, used to normalize the parameter scales.
The integrals in the likelihood are computed on a $128^2 \times 1024$ grid over $(v_1, v_2, E_\nu)$.

All metrics we have examined paint a similar picture.
Increasing model size generally improves the quality of the model, especially $H$.
While not apparent from KL divergences, the oscillation parameter metric indicates that this improvement has not saturated even at the largest $H=128$ examined (larger $H$ were infeasible on a single A100 GPU for the range of $d$ considered).
Increasing $n_\mathrm{layers}$ has a relatively small effect except at small $H$, but gives smoother behavior in $H$; we have not identified the precise source of non-smooth variation in the curves, but it can result from a combination of seed dependence, precise choice of stopping condition for training, the interaction of the model size with grid resolution, and various other factors.
Notably, dependence on the grid resolution $d$ is weak, with no pattern apparent.
This suggests that training is unlikely to be data-limited.

These results indicate that hyperparameter dependence has weak effects on the results once a threshold model size and resolution is achieved.
It is likely that this behavior can be understood in terms of the infinite-size properties of neural parameterizations, including universal function approximation and the emergence of Gaussian process statistics, but further investigation will be required to develop a more precise and quantitative understanding.
Qualitatively, however, it suggests that good results can be obtained simply by scaling up.

Variation over architecture and training hyperparameters could be somehow incorporated into systematic uncertainty estimation, but there is an infinite space of possibilities and no principled reason for including and excluding particular variations. 
The wisdom and efficacy of various constructions can only be evaluated in detailed calibration tests.
Since the simple scheme of estimating ML systematic uncertainties through random network initialization (in conjunction with variations describing experimental systematics) is sufficient to provide a reasonably well-calibrated scheme in the tests performed here, explicit studies of other possibilities are left for future work.

\bibliographystyle{apsrev4-2}
\bibliography{refs}

\end{document}